\documentclass[twocolumn,showpacs,nofootinbib,preprintnumbers,amsmath,amssymb,showkeys,superscriptaddress]{revtex4-1}
\usepackage{graphicx}
\usepackage{amsmath}
\usepackage{amssymb}
\usepackage{dcolumn}
\usepackage{bm}
\usepackage{xcolor}
\usepackage{dsfont}
\usepackage{feynmp}
\usepackage{marvosym}
\usepackage{phaistos}
\usepackage[hidelinks=true]{hyperref}

\newcommand \beq{\begin{eqnarray}}
\newcommand \eeq{\end{eqnarray}}

\begin{document}
\unitlength=1mm
\allowdisplaybreaks

\title{The infrared-safe Minkowskian Curci-Ferrari model}

\author{Santiago Oribe}
\affiliation{Instituto de F\'isica, Facultad de Ingenier\'ia, Universidad de la Rep\'ublica, Montevideo, Uruguay.}
\affiliation{Centre de Physique Th\'eorique, CNRS, École Polytechnique, IP Paris, F-91128 Palaiseau, France.}

\author{Marcela Pel\'aez}
\affiliation{Instituto de F\'isica, Facultad de Ingenier\'ia, Universidad de la Rep\'ublica, Montevideo, Uruguay.}

\author{Urko Reinosa}
\affiliation{Centre de Physique Th\'eorique, CNRS, École Polytechnique, IP Paris, F-91128 Palaiseau, France.}

\date{\today}

\begin{abstract}
We discuss the existence of Landau-pole-free renormalization group trajectories in the Minkowskian version of the Curci-Ferrari model as a function of a running parameter $q^2$ associated to the four-vector $q$ at which renormalization conditions are imposed, and which can take both space-like ($\smash{q^2<0}$) and time-like ($\smash{q^2>0}$) values. We discuss two possible extensions of the infrared-safe scheme defined in Ref.~\cite{Tissier:2011ey} for the Euclidean version of the model, which coincide with the latter in the space-like region upon identifying $\smash{Q^2\equiv-q^2}$ with the square of the renormalization scale in that reference. The first extension uses real-valued renormalization factors and leads to a flow in the time-like region with a similar structure as the flow in the space-like region (or in the Euclidean model), including a non-trivial fixed point and a family of trajectories bounded at all scales by the value of the coupling at this fixed point. Interestingly, the fixed point in the time-like region has a much smaller value of $\lambda\equiv g^2N/16\pi^2$ than the corresponding one in the space-like region, a value closer to the perturbative boundary $\smash{\lambda=1}$. However, in this real-valued infrared-safe scheme, the flow cannot connect the time-like and space-like regions. Thus, it is not possible to deduce the relevant time-like flow trajectory from the sole knowledge of a space-like flow trajectory. To try to cure this problem, we investigate a second extension of the Euclidean IR-safe scheme, which allows for complex-valued renormalization factors. We discuss under which conditions these schemes can make sense and study their ability to connect space- and time-like flow trajectories. In particular, we investigate to which types of time-like trajectories the perturbative space-like trajectories are mapped onto.
\end{abstract}

\maketitle

\section{Introduction}
  
The Curci-Ferrari model was introduced in the mid-70's as an example of a renormalizable theory involving massive gauge fields \cite{Curci:1976bt}. Although abandoned due to its perturbative nonunitarity,\footnote{The question of unitarity is still opened as a real proof would need to take into consideration the fact that the asymptotic states in this model may not be the elementary fields appearing in the Lagrangian.} it has recently been revived as a phenomenological way to take into account the dynamical generation of mass observed in Landau-gauge lattice simulations of Yang-Mills theories, while extending the incomplete Faddeev-Popov gauge-fixed action in the continuum formulation of the Landau gauge \cite{Tissier:2010ts,Tissier:2011ey}. 

Despite its phenomenological nature, the model has shown a rather impressive capability at reproducing many established features of Landau-gauge Yang-Mills theories, such as decoupling or positivity violation in two- and three-point correlation functions \cite{Pelaez:2013cpa,Pelaez:2014mxa,Pelaez:2015tba,Figueroa:2021sjm,Gracey:2019xom,Barrios:2020ubx,Barrios:2021cks,Barrios:2022hzr,Barrios:2024ixj}, while giving access to observables such as the YM confinement/deconfinement transition temperature or the associated order parameter known as the Polyakov loop \cite{Reinosa:2014zta,Reinosa:2015gxn,Reinosa:2016xaj,Maelger:2018vow,Surkau:2024zfb}. It has also been extended to QCD where it properly captures the spontaneous breaking of chiral symmetry and the associated generation of mass \cite{Pelaez:2017bhh,Pelaez:2020ups} while giving access to observables such as the pion decay constant \cite{Pelaez:2022rwx} or the energy of a quark-antiquark pair \cite{Pelaez:2024mtq}. 

Overall, the to date established predictions of the model are in good agreement with non-perturbative approaches to QCD, such as lattice simulations when available, and continuum methods such as Schwinger-Dyson equations (see e.g. \cite{Roberts:1994dr,Alkofer:2000wg,Fischer:2006ub,Roberts:2007ji,Binosi:2009qm,Cloet:2013jya,Aguilar:2015bud,Huber:2018ned,Papavassiliou:2022wrb,Binosi:2014aea,Maris:1997tm,Maris:2003vk,Eichmann:2008ef,Cloet:2008re,Boucaud:2008ky,Eichmann:2009qa,Fischer:2008uz,Boucaud:2008ky,Rodriguez-Quintero:2010qad,Pennington:2011xs,Huber:2012zj,Cloet:2013jya,Boucaud:2008ky,Roberts:2020hiw,Roberts:2020hiw,Gao:2021wun,Huber:2016tvc,Huber:2020keu,Aguilar:2023mam,Aguilar:2023qqd,Aguilar:2024fen}) or the functional renormalization group (see e.g. \cite{Cyrol:2014kca,Braun:2007bx,Fister:2013bh,Pawlowski:2003hq,Pawlowski:2005xe,Cyrol:2017ewj,Cyrol:2018xeq,Corell:2018yil,Blaizot:2021ikl,Horak:2021pfr}).

The success of the model is rooted in the existence of Landau-pole-free renormalization group trajectories along which the coupling in the glue sector (or rather the perturbative expansion parameter, see below) remains moderate \cite{Reinosa:2017qtf,Gracey:2019xom}. These infrared-safe (IR-safe) trajectories allow for the use of perturbative schemes\footnote{Other perturbative approaches include the screened massive expansion \cite{Siringo:2015wtx,Siringo:2019qwx,Comitini:2020ozt,Comitini:2021kxj}, and the refined Gribov-Zwanziger approach \cite{Mintz:2017qri,Barrios:2024idr,deBrito:2024ffa}.} in applications of the model to pure YM and heavy-quark QCD theories \cite{Reinosa:2017qtf} and of semi-perturbative schemes in applications to QCD \cite{Pelaez:2017bhh,Pelaez:2020ups}. 

To date, however, although correlators have been evaluated both in the Euclidean and in the Minkowskian formulations of the model,\footnote{See Ref.~\cite{Hayashi:2020few} for this latter case.} the evaluation of observables has been restricted to those that can be reached within the Euclidean formulation. For instance, the pion decay constant was computed in the chiral limit as this requires knowing the correlation functions for vanishing Euclidean momenta.\footnote{The physical value can then be deduced using chiral perturbation theory.} Similarly, the study of the phase diagram can be entirely done within the Euclidean formulation.


Some other observables, such as the spectrum of hadrons require to work in a Minkowskian formulation. It is then important to ask whether the properties of the Euclidean Curci-Ferrari model extend to its Minkowskian version. In particular, what is the nature of the renormalization group flow in this case? Are there Landau-pole free trajectories? Are the perturbative Euclidean flows mapped into perturbative Minkowskian flows? In this work, we investigate these questions at one-loop order in the pure YM case.

After a recap of the Euclidean CF model, we explain how to obtain the one-loop Minkowskian self-energies without effort, by exploiting complex analysis. We then define two schemes that extend the IR-safe scheme of the Euclidean model. The first one uses real-valued renormalization factors and leads to IR-safe trajectories both in the space-like region (which coincides with the Euclidean region) and in the time-like region. These two regions are not connected by the flow, however. We then consider a complex-valued renormalization scheme which allows one to connect the two regions and we investigate which perturbative space-like trajectories remain perturbative in the time-like region.

Both schemes that we consider belong to the class of Taylor schemes whose running coupling is intimately related to the gluon and ghost two-point functions. For the sake of brievety, however, we shall postpone the study of the latter to a subsequent work.


\section{Recap of the Euclidean CF model}
The Euclidean Curci-Ferrari (CF) model is defined as
\beq
\hat S & = & \int d^d\hat x\,\,\Bigg\{\,\frac{1}{4}\hat F_{\mu\nu}^a\hat F_{\mu\nu}^a+\frac{1}{2}m^2\hat A_\mu^a\hat A_\mu^a\nonumber\\
& & \hspace{1.5cm} +\,\hat\partial_\mu\hat{\bar{c}}^a\hat D_\mu \hat c^a+i\hat h^a\hat\partial_\mu \hat A_\mu^a\Bigg\}\,,\label{eq:Stilde}
\eeq
with $\smash{\hat F_{\mu\nu}^a\equiv\hat\partial_\mu\hat A_\nu^a-\hat\partial_\nu\hat A_\mu^a+gf^{abc}\hat A_\mu^b\hat A_\nu^c}$ the Euclidean version of the field-strength tensor and $\hat D_\mu\varphi^a\equiv \hat\partial_\mu\varphi^a+gf^{abc}\hat A_\mu^b\varphi^c$ the adjoint covariant derivative.  Here and in what follows, the hat refers to Euclidean quantities. 

\subsection{Feynman rules}
Although we shall not be repeating here the Euclidean calculations of Ref.~\cite{Tissier:2011ey}, it will be useful to specify our choice of conventions for the Feynman rules. The comparison with the corresponding conventions in the Minkowskian model will allow us to deduce, without effort, the Minkowskian results from the Euclidean ones.

In the Euclidean, we take the Fourier convention $\smash{\hat\partial_\mu\to -iK_\mu}$, with upper case letters denoting Euclidean momenta such that $\smash{K^2=K_\mu K_\mu}$. We also make the choice to include the minus sign that comes from $e^{-\hat S}$ in front of each of the tree-level vertices. In this way, the number of minus signs to be dealt with when applying the Feynman rules is minimized.\footnote{Of course, one should not forget that each ghost loop yields an extra sign.} 

The tree-level ghost and gluon propagators read, respectively,
\beq
\frac{\delta^{ab}}{P^2} \quad {\rm and} \quad \delta^{ab}\frac{\hat P^\perp_{\mu\nu}(P)}{P^2+m^2}\,,
\eeq
with $\smash{\hat P^\perp_{\mu\nu}(P)\equiv \delta_{\mu\nu}-P_\mu P_\nu/P^2}$ the transversal projector with respect to momentum $P$ in Euclidean metric. As for the tree-level vertices, we have
\beq
igf^{abc}P_\mu\,,
\eeq
for the ghost-antighost-gluon vertex, where $a$, $b$ and $c$ are the colors carried by the antighost, gluon and ghost respectively and $P$ is the momentum carried by the outgoing antighost,
\beq
\frac{ig}{3!}f^{abc} \Big[(P-R)_\nu\delta_{\mu\rho}+(R-Q)_\mu\delta_{\rho\nu}+(Q-P)_\rho\delta_{\nu\mu}\Big]\,,\label{eq:v3}\nonumber\\
\eeq
for the three-gluon vertex, where $(a,P,\mu)$, $(b,Q,\nu)$ and $(c,R,\rho)$ are associated to each leg and the momenta are outgoing, and finally
\beq
& & -\frac{g^2}{4!}\Big[f^{abe}f^{cde}(\delta_{\mu\rho}\delta_{\nu\sigma}-\delta_{\mu\sigma}\delta_{\nu\rho})\nonumber\\
& & \hspace{1.5cm}+\,f^{ace}f^{dbe}(\delta_{\mu\sigma}\delta_{\rho\nu}-\delta_{\mu\nu}\delta_{\rho\sigma})\nonumber\\
& & \hspace{2.5cm}+\,f^{ade}f^{bce}(\delta_{\mu\nu}\delta_{\sigma\rho}-\delta_{\mu\rho}\delta_{\sigma\nu})\Big]\,,
\eeq
for the four-gluon vertex. 

\subsection{One-loop self-energies}
The exact ghost and gluon propagators have a similar structure than the tree-level ones. They write
\beq
\delta^{ab} \hat G_c(P^2) \quad {\rm and} \quad \delta^{ab}\hat P^\perp_{\mu\nu}(P)\hat G_A(P^2)\,,\label{eq:props}
\eeq
with 
\beq
\hat G_c^{-1}(P^2) & = & P^2+\hat \Pi_c(P^2)\,,\\
\hat G_A^{-1}(P^2) & = & P^2+m^2+\hat \Pi_A(P^2)\,,
\eeq
and where the ghost and gluon self-energies $\hat\Pi_c(P^2)$ and $\hat\Pi_A(P^2)$ are given by $-1$ times the sum of the corresponding one-particle-irreducible (1PI) diagrams.\footnote{Our convention for a self-energy is always that it represents the correction to the square mass, when any, and this, both in the Euclidean and in the Minkowskian cases.} 

At one-loop order, $\hat\Pi_c(P^2)$ and $\hat\Pi_A(P^2)$ can be written as \cite{Passarino:1978jh}:
\beq
\hat\Pi(P^2) & = & \alpha_m\left(\frac{P^2}{m^2}\right)\,\hat A_m\nonumber\\
& + & m^2\!\!\!\!\sum_{m_i\in\{0,m\}}\!\!\!\!\beta_{m_1m_2}\left(\frac{P^2}{m^2}\right)\,\hat B_{m_1m_2}(P^2)\,,\label{eq:reduc}
\eeq
where the Euclidean tadpole and bubble master integrals are defined as
\beq
\hat A_m & \equiv & \int_Q\frac{1}{Q^2+m^2}\,,\label{eq:AE}\\
\hat B_{m_1m_2}(P^2) & \equiv & \int_Q\frac{1}{Q^2+m^2_1}\frac{1}{(Q+P)^2+m^2_2}\,,\label{eq:BE}
\eeq
with
\beq
\int_Q\equiv \Lambda^{4-d}\int\frac{d^dQ}{(2\pi)^d}\,,
\eeq
and $\Lambda$ the scale associated to dimensional regularization.\footnote{In dimensional regularization, the coupling constant has mass dimension $(4-d)/2$ and it customary to factor out this dimension explicitely as $\Lambda^{(4-d)/2}$.} The benefit of the master integrals is of course that they possess well known explicit expressions (which we will not recall here) in an expansion around $\smash{d=4}$.

The coefficients $\alpha_m$ and  $\beta_{m_1m_2}$ of the decomposition (\ref{eq:reduc}) are dimensionless rational fractions of $P^2/m^2$ whose expressions for each self-energy we recall in App.~\ref{app:coeff}.\footnote{Some of these coefficients can also depend on $d$ but we leave this dependence implicit.}

\subsection{Infrared-safe renormalization factors}
At one-loop order, the renormalized propagators take a similar form as in Eq.~(\ref{eq:props}), with
\beq
& & \hat G_c^{-1}(P^2;Q^2)=\hat Z_c(Q^2)P^2\!+\!\hat \Pi_c(P^2)\,,\\
& & \hat G_A^{-1}(P^2;Q^2)\nonumber\\
& & \hspace{0.2cm}=\,\hat Z_A(Q^2)P^2\!+\!\hat Z_A(Q^2)\hat Z_{m^2}(Q^2)m^2\!+\!\hat \Pi_A(P^2)\,,\label{eq:ccc}
\eeq
and where, for reasons that will become clear below, we have denoted the renormalization scale as $Q^2$. 

The renormalization factors $\hat Z_X$ with $\smash{X\in\{c,A,m^2\}}$ are needed to absorb the UV divergences that are present in the self-energies. Their exact expressions depend on the chosen renormalization scheme. In the infrared-safe (IR-safe) scheme, they are fixed from the conditions \cite{Tissier:2011ey}
\beq
\hat G_c^{-1}(P^2=Q^2;Q^2) & = & Q^2\,,\label{eq:rc1}\\
\hat G_A^{-1}(P^2=Q^2;Q^2) & = & Q^2+m^2\,,\label{eq:rc2}
\eeq
and
\beq
\hat Z_A\hat Z_c\hat Z_{m^2}=1\,.\label{eq:Zm}
\eeq
Note that this last condition is only possible because the combination of renormalization factors appearing in the left-hand side is finite, owing to one of the two non-renormalization theorems of the CF model \cite{Gracey:2002yt,Dudal:2002pq,Wschebor:2007vh,Tissier:2008nw}. The IR-safe scheme includes a last condition
\beq
\hat Z_A\hat Z_c^2\hat Z_{g^2}=1\,,\label{eq:Zg}
\eeq
which fixes the renormalization of the coupling in terms of the other renormalization factors. Again, this is only possible because of a second non-renormalization theorem, known as Taylor's theorem \cite{Taylor:1971ff}. For a recent review of the two non-renormalization theorems, see Ref.~\cite{Reinosa:2024vph}.

The IR-safe scheme is peculiar since the renormalization factors for $g^2$ and $m^2$ are fixed solely in terms of $\hat Z_A$ and $\hat Z_c$. To determine the latter one first notices that, to one-loop accuracy,
\beq
\hat Z_A\hat Z_{m^2}=\hat Z_c^{-1}\simeq 2-\hat Z_c\,.
\eeq
Upon plugging this back into Eq.~(\ref{eq:ccc}), the set of equations (\ref{eq:rc1})-(\ref{eq:rc2}) turns into a triangular linear system for $\hat Z_c$ and $\hat Z_A$, from which one deduces that
\beq
\hat Z_c(Q^2) & = & 1-\frac{\hat \Pi_c(Q^2)}{Q^2}\,,\label{eq:Zc}\\
\hat Z_A(Q^2) & = & 1-m^2\frac{\hat\Pi_c(Q^2)}{(Q^2)^2}-\frac{\hat\Pi_A(Q^2)}{Q^2}\,.\label{eq:ZA}
\eeq
Note that the explicit poles in these expressions are spurious because both $\hat\Pi_c(Q^2)$ and $\hat\Pi_A(Q^2)+(m^2/Q^2)\hat\Pi_c(Q^2)$ are proportional to $Q^2$ as $\smash{Q^2\to 0}$. The first result follows from the derivative nature of the ghost-antighost-gluon vertex, the transversality of the gluon propagator and the presence of a gluon mass that regulates the infrared, see Ref.~\cite{Barrios:2020ubx}. The second result follows from a symmetry constraint relating the ghost self-energy and the longitudinal component $\hat\Pi_\parallel(Q^2)$ of the gluon self-energy \cite{Tissier:2011ey}. At one-loop order, this identity writes
\beq
\hat\Pi_\parallel(Q^2)+(m^2/Q^2)\hat\Pi_c(Q^2)=0\,.\nonumber
\eeq 
Owing to the fact that $\smash{\hat\Pi_\parallel(0)=\hat\Pi_A(0)}$, this implies that $\hat\Pi_A(Q^2)+(m^2/Q^2)\hat\Pi_c(Q^2)$ vanishes in the limit $\smash{Q^2\to 0}$, and, due to the presence of the gluon mass that regulates the infrared, that $\hat\Pi_A(Q^2)+(m^2/Q^2)\hat\Pi_c(Q^2)$ is proportional to $Q^2$ in this limit.

\begin{center}
\begin{figure}[t]
\includegraphics[height=0.3\textheight]{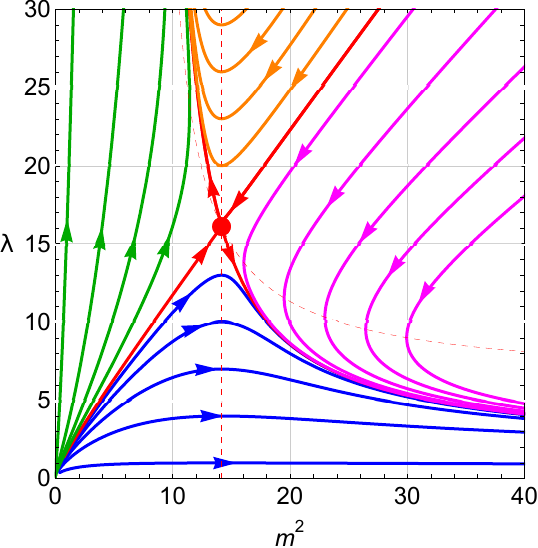}
\caption{One-loop flow of the Euclidean Curci-Ferrari model in the IR-safe scheme, in the plane $(\hat m^2,\lambda)$. The dashed lines correspond to points at which one of the two beta functions vanishes. They are useful for the classification of the various flow trajectories, see the Appendix. The arrows show the direction of the flow from UV to IR scales.}
\label{fig:diag}
\end{figure}
\end{center}

\vglue-6mm

\subsection{Anomalous dimensions and beta functions}
From the renormalization factors $\hat Z_X$, one can deduce the associated anomalous dimensions\footnote{Our choice of writing $2Q^2d/dQ^2$ comes from the fact that, in the Minkowskian, $Q^2$ will be replaced by $q^2$ which can take both positive and negative values.}
\beq
\hat \gamma_X(Q^2)\equiv 2 Q^2\frac{d\ln \hat Z_X}{dQ^2}\,,
\eeq
and then the relevant beta functions:
\beq
\hat \beta_{g^2}(Q^2) & \equiv & 2Q^2\frac{dg^2}{dQ^2}=-g^2\hat\gamma_{g^2}(Q^2)\,,\\
\hat \beta_{m^2}(Q^2) & \equiv & 2Q^2\frac{dm^2}{dQ^2}=-m^2\hat\gamma_{m^2}(Q^2)\,.
\eeq
One very convenient feature of the IR-safe scheme is that, from Eqs.~(\ref{eq:Zm}) and (\ref{eq:Zg}), the various anomalous dimensions are related by $\smash{\hat\gamma_A+\hat\gamma_c+\hat\gamma_{m^2}=0}$ and $\smash{\hat\gamma_A+2\hat\gamma_c+\hat\gamma_{g^2}=0}$ and thus the beta functions are fully determined once the field anomalous dimensions are known:
\beq
\hat\beta_{g^2} & = & g^2\left(\hat\gamma_A+2\hat\gamma_c\right),\\
\hat\beta_{m^2} & = & m^2\left(\hat\gamma_A+\hat\gamma_c\right).
\eeq
It is in fact more convenient to work with the variables $\smash{\lambda\equiv g^2N/16\pi^2}$ and $\smash{\hat m^2\equiv m^2/Q^2}$ whose corresponding $\beta$-functions are
\beq
\hat\beta_\lambda & = & \lambda\left(\hat\gamma_A+2\hat\gamma_c\right),\\
\hat\beta_{\hat m^2} & = & \hat m^2\left(\hat\gamma_A+\hat\gamma_c-2\right).
\eeq
Working with dimensionless parameters is mandatory if one wants to study the renormalization group fixed points, see below. The rationale for introducing $\lambda$ is that this is the actual expansion parameter of perturbative calculations within the CF model. As long as this parameter remains far below $1$, the one-loop calculations used in this scheme are under control.

\subsection{Renormalization group flow}
At one-loop order, $\hat\gamma_A$ and $\hat\gamma_c$ can be derived from Eqs.~(\ref{eq:Zc})-(\ref{eq:ZA}). They are found to be \cite{Tissier:2011ey}
\beq
\hat\gamma_c(Q^2) & = & -\frac{\lambda}{2T^2} \Big[2T^2+2T -T^3 \ln T\nonumber\\
& & \hspace{1.0cm} +\,(T+1)^2 (T-2) \ln (T+1)\Big]\,,\label{eq:ggc}
\eeq
and
\beq
\hat\gamma_A(Q^2) & = & \frac{\lambda}{6T^3}  \Bigg[-17 T^3+74T^2-12T+T^5 \ln T\nonumber\\
& & \hspace{1.0cm}-\,(T-2)^2 (2T-3) (T+1)^2 \ln (T+1) \nonumber\\\nonumber\\
& & \hspace{1.0cm}-\,T^{3\over2} \sqrt{T+4} \left(T^3-9T^2+20T-36\right)\nonumber\\
& & \hspace{1.5cm}\times\,\ln\!\left(\frac{\sqrt{T+4}-\sqrt{T}}{\sqrt{T+4}+\sqrt{T}}\right)\!\!\Bigg]\,,\label{eq:gga}
\eeq
with $\smash{T\equiv Q^2/m^2=1/\hat m^2}$. The resulting Euclidean flow diagram is shown in Fig.~\ref{fig:diag}. As discussed in App.~\ref{app:diagram}, it describes four types of trajectories, and, therefore, four types of theories:
\begin{itemize}

\item[$\bullet$] {\bf Type 1:} theories that are asymptotically free both in the UV and in the IR (blue trajectories in Fig.~\ref{fig:diag}). For these theories, the parameter $\hat m^2$ goes to $0$ in the UV and to $\infty$ in the IR;\footnote{However, the mass parameter $m^2$ goes to $0$ both in the UV and in the IR, see Refs.~\cite{Tissier:2011ey,Reinosa:2017qtf}.}

\item[$\bullet$] {\bf Type 2:} theories that are asymptotically free in the UV and possess an IR Landau pole (green trajectories). For these theories, the parameter $\hat m^2$ goes to $0$ in the UV and to a finite value $\hat m^2_{\rm LP}$ as one approaches the IR Landau pole;

\item[$\bullet$] {\bf Type 3:} theories that are asymptotically free in the IR and possess an UV Landau pole (purple trajectories). For these theories, the parameter $\hat m^2$ goes to infinity in the IR and also as one approaches the UV Landau pole;

\item[$\bullet$] {\bf Type 4:} theories that possess both an IR and an UV Landau pole (orange trajectories). For these theories, the parameter $\hat m^2$ approaches $\hat m^2_{\rm LP}$ as one approches the IR Landau pole, and diverges as one approaches the UV Landau pole.

\end{itemize}
 
\begin{center}
\begin{figure}[t]
\includegraphics[height=0.3\textheight]{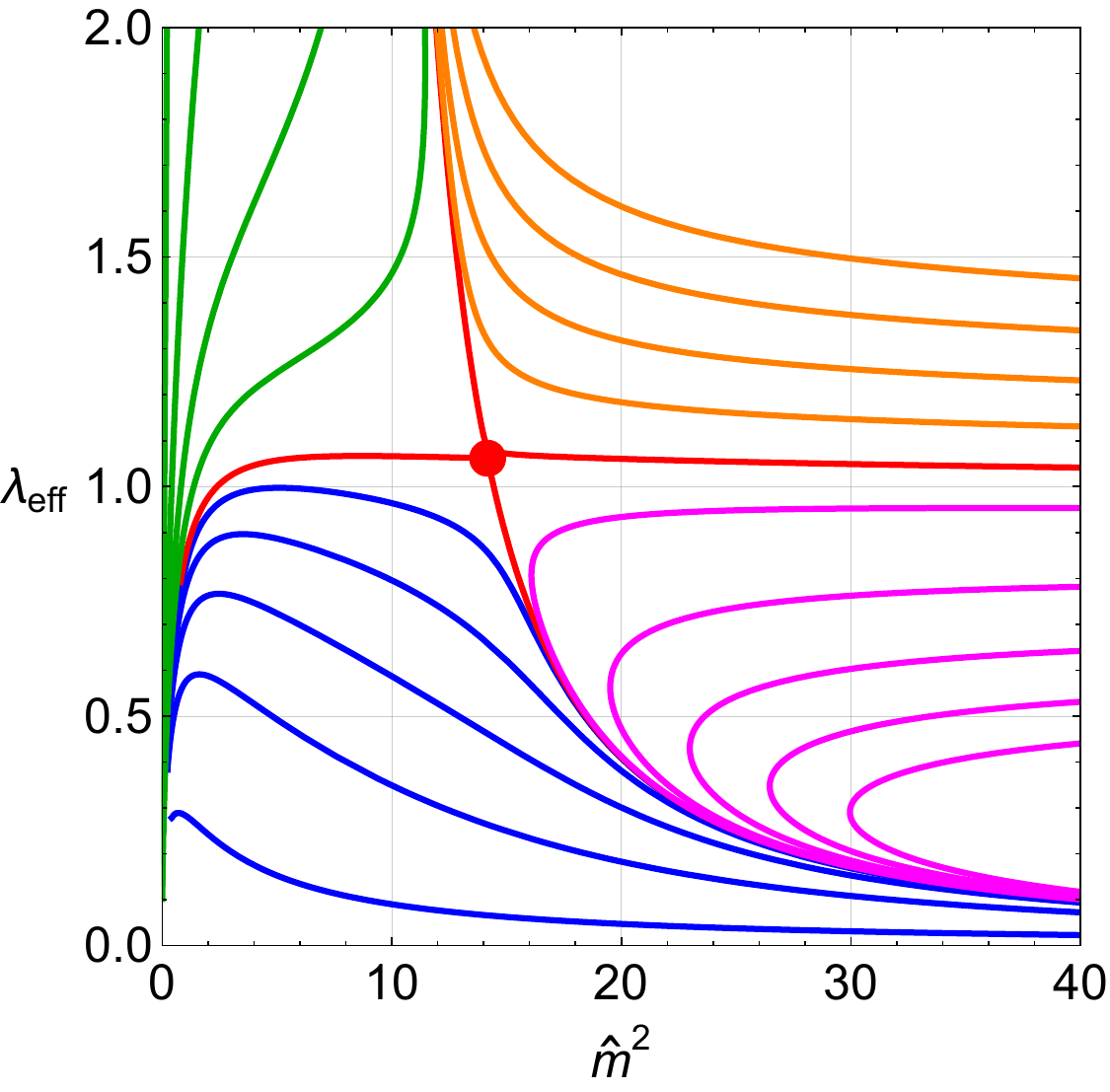}
\caption{One-loop flow of the Euclidean Curci-Ferrari model in the IR-safe scheme, in the plane $(\hat m^2,\lambda_{\rm eff})$. For the direction of the flow, see Fig.~\ref{fig:diag}.}
\label{fig:diag_tilde}
\end{figure}
\end{center}

The flow diagram possesses two fixed points at which both beta functions $\beta_\lambda$ and $\beta_{\hat m^2}$ vanish \cite{Reinosa:2017qtf}. In addition to the fully attractive UV fixed point at $(0,0)$, there is a non-trivial fixed point obtained by noticing that $\hat\beta_\lambda$ takes the form $\smash{\lambda^2 \hat{f}(\hat m^2)}$ with $\hat f(\hat m^2)$ vanishing at $\smash{\hat m^2=\hat m^2_\star\simeq 14.17}$. The beta function thus vanishes at this value of $\hat m^2$. On the other hand, the structure of $\hat\beta_{\hat m^2}$ is $\hat m^2(\lambda \hat h(\hat m^2)-2)$ which also vanishes at $\hat m^2=\hat m^2_\star$ provided one chooses $\lambda=\hat\lambda_\star\equiv 2/\hat h(\hat m^2_\star)\simeq 16.11$. We have thus a non-trivial fixed point at $(\hat m^2_\star,\hat\lambda_\star)\simeq(14.17,16.11)$. The value of $\hat\lambda_\star$ is quite high, certainly way above the regime of validity of the one-loop calculation, see however the remark in the next section.

We mention that $\smash{(\hat m^2,\lambda)=(\infty,0)}$ is also a fixed point in the sense that $\beta_\lambda$ and $\beta_{1/\hat m^2}$ vanish at this point. However, there are no fixed points with $\smash{\lambda=\infty}$ because $\smash{1/\lambda=0}$ is always reached at a finite scale (Landau pole). Another way to see this is to notice that the beta function for $1/\lambda$,
\beq
\hat\beta_{1/\lambda}=-\hat f(\hat m^2)\,,
\eeq
cannot vanish away from $\smash{\hat m^2=\hat m^2_\star}$, see App.~\ref{app:diagram} for more details.

\subsection{Effective coupling}

It has been argued in Ref.~\cite{Tissier:2011ey} that an even more sensible measure of the validity of perturbation theory within the CF model is $\lambda_{\rm eff}\equiv\lambda/(1+\hat m^2)$.  The flow diagram in the plane $(\hat m^2,\lambda_{\rm eff})$ is represented in Fig.~\ref{fig:diag_tilde}. At the non-trivial fixed point, the effective coupling is $\lambda_{\rm eff}^\star\simeq 1.06$.

Interestingly enough, the effective coupling can be seen as the coupling in a renormalization scheme obtained from (\ref{eq:rc1})-(\ref{eq:Zg}) upon replacing (\ref{eq:rc2}) by
\beq
\hat G_A^{-1}(P^2=Q^2;Q^2)=Q^2\,.\label{eq:rc3}
\eeq
This condition, together with (\ref{eq:rc1}) and (\ref{eq:Zg}), is more in line with the one used in lattice simulations. Of course, lattice simulations do not require (\ref{eq:Zm}). This is an added feature of the model because we have one extra parameter to fix.

While the ghost renormalization factor $Z_c(Q^2)$ is not changed, the gluon renormalization factor $Z_A(Q^2)$ is changed to
\beq
Z'_A(Q^2)=\frac{Q^2}{Q^2+m^2}Z_A(Q^2)\,.
\eeq
and owing to (\ref{eq:Zm}) and (\ref{eq:Zg}), this leads to
\beq
Z'_{m^2}(Q^2) & = & \frac{Q^2+m^2}{Q^2}Z_{m^2}(Q^2)\,,\\
Z'_{g^2}(Q^2) & = & \frac{Q^2+m^2}{Q^2}Z_{g^2}(Q^2)\,.
\eeq
Since the bare coupling should not depend on the scheme, we deduce that the coupling constants in the new and old scheme are related by
\beq
\lambda'(Q^2)=\frac{Q^2}{Q^2+m^2}\lambda(Q^2)\,,
\eeq
and thus that $\lambda_{\rm eff}(Q^2)$ coincides with $\lambda'(Q^2)$, as announced.


\section{The Minkowskian CF model}
To associate a Minkowskian model to the Euclidean CF model (\ref{eq:Stilde}), we perform the formal replacements:
\beq
\hat x_0\to ix^0=ix_0\,, \quad \hat x_i\to x^i=-x_i\,,
\eeq
\beq
\hat\partial_0\to -i\partial_0=-i\partial^0\,, \quad \hat\partial_i\to \partial_i=-\partial^i\,,
\eeq
as well as 
\beq
\hat A_0^a\to -iA_0^a=-iA^{a,0}\,, \quad \hat A_i^a\to A^{a,i}=-A_i^a\,.
\eeq
Although it would be possible to leave the ghost, antighost and Nakanishi-Lautrup fields unchanged, it is also convenient to consider the replacements
\beq
\hat{c}^a\to i c^a\,, \quad \hat{\bar{c}}^a\to i\bar c^a\,, \quad\hat h^a\to i h^a\,.
\eeq
The Minkowskian model $S$ is then defined from $-\hat S\to iS$. In the present case, it is easily verified that
\beq
S & = & \int d^dx\,\Bigg\{\!\!-\frac{1}{4}F_{\mu\nu}^aF^{\mu\nu,a}+\frac{1}{2}m^2A_\mu^aA^{\mu,a}\nonumber\\
& & \hspace{1.5cm}-\,\partial_\mu\bar c^a D^\mu c^a-h^a \partial_\mu A^{\mu,a}\Bigg\},
\eeq
with $F_{\mu\nu}^a=\partial_\mu A_\nu^a-\partial_\nu A_\mu^a+gf^{abc}A_\mu^b A_\nu^c$ and $D_\mu\varphi^a=\partial_\mu\varphi^a+gf^{abc}A_\mu^b\varphi^c$.

\subsection{Feynman rules}
As before, our convention for the Feynman rules is to absorb the factor of $i$ that comes from $e^{iS}$ in front of each of the tree-level vertices. Moreover, our Fourier convention is now $\partial_\mu\to ik_\mu$, where lower case letters denote Minkowkian momenta such that $k^2=k^\mu k_\mu$. With these conventions, the Feynman rules are
\beq
\frac{-i\delta^{ab}}{p^2+i0^+} \quad {\rm and} \quad \frac{-i\delta^{ab}}{p^2-m^2+i0^+}P^\perp_{\mu\nu}(p)
\,,
\eeq
for the ghost and gluon propagators, with $P^\perp_{\mu\nu}(p)\equiv g_{\mu\nu}-p_\mu p_\mu/p^2$ the transerval projector with respect to momentum $p$ in Minkowskian metric,
\beq
gf^{abc}p_\mu\,,
\eeq
for the ghost-antighost-gluon vertex, where $a$, $b$ and $c$ are the colors carried by the antighost, gluon and ghost respectively and $p$ is the momentum carried by the outgoing antighost,
\beq
\frac{g}{3!}f^{abc} \Big[(p-r)_\nu g_{\mu\rho}+(r-q)_\mu g_{\rho\nu}+(q-p)_\rho g_{\nu\mu}\Big],\label{eq:v3}
\eeq
for the three-gluon vertex, where $(a,p,\mu)$, $(b,q,\nu)$ and $(c,r,\rho)$ are associated to each leg and the momenta are ougoing, and finally
\beq
& & -i\frac{g^2}{4!}\Bigg[f^{abe}f^{cde}( g_{\mu\rho} g_{\nu\sigma}- g_{\mu\sigma} g_{\nu\rho})\nonumber\\
& & \hspace{1.4cm}+\,f^{ace}f^{dbe}( g_{\mu\sigma} g_{\rho\nu}- g_{\mu\nu} g_{\rho\sigma})\nonumber\\
& & \hspace{2.0cm}+\,f^{ade}f^{bce}( g_{\mu\nu} g_{\sigma\rho}- g_{\mu\rho} g_{\sigma\nu})\Bigg],
\eeq
for the four-gluon vertex. 

Our choices are such that, with respect to the Euclidean Feynman rules, and in addition to the obvious replacements $P^2\to p^2$, $m^2\to -m^2+i0^+$ and $\delta_{\mu\nu}\to g_{\mu\nu}$, the propagators both receive an extra factor of $-i$. The same goes for the three-point vertices, whereas the four-point vertex receives an extra factor of $i$. This will come in handy when relating the diagrams in the Euclidean and in the Minkowskian CF models.

\subsection{One-loop self-energies}\label{sec:1loop_M}
The full propagators now take the form
\beq
-i\delta^{ab} G_c(p^2) \quad {\rm and} \quad -\!i\delta^{ab}P^\perp_{\mu\nu}(p) G_A(p^2)\,,
\eeq
with
\beq
G_c^{-1}(p^2) & = & p^2+i0^+-\Pi_c(p^2)\,,\\
G_A(p^2) & = & p^2-m^2+i0^+-\Pi_A(p^2)\,.
\eeq
The self-energies $\Pi_c(p^2)$ and $\Pi_A(p^2)$ are this time given by $-i$ times the sum of corresponding 1PI diagrams.  

At one-loop order, one could of course compute the corresponding Feynman diagrams directly in Minkowskian. There is a simpler way, however, that exploits the explicit Euclidean result for the self-energies. In fact, for space-like momenta $\smash{p^2<0}$, the Minkowskian self-energies are directly connected to the Euclidean self-energies as
\beq
\Pi(p^2<0)=\hat\Pi(-p^2)\,,\label{eq:c0}
\eeq
and, more generally, for any Minkowskian momentum $\smash{p^2\in\mathds{R}}$ (be it space-like or time-like), the Minkowskian self-energies can be obtained as
\beq
\Pi(p^2\in\mathds{R})=\bar\Pi(z=p^2+i0^+)\,,\label{eq:c1}
\eeq
where $\bar\Pi(z)$ is the unique analytic function over $\mathds{C}/\mathds{R}_+$ that connects with the Euclidean result over $\mathds{R}_-^*$ as
\beq
\bar\Pi(z=p^2<0)=\hat\Pi(-p^2)\,.\label{eq:c2}
\eeq
That the continuation is unique follows from a well known result in complex analysis. The existence of the continuation is more subtle specially in theories with positivity violation where the standard K\"allen-Lehmann spectral representation does not hold. We will show below that, for the one-loop self-energies of the CF model, the continuation actually exists. Knowing that a continuation exists and that it is unique will then allows us to construct the continuation from the sole knowledge of the Euclidean self-energies, and then, to deduce Minkowskian self-energies without effort, using Eq.~(\ref{eq:c1}). 

Let us note that, by combining Eqs.~(\ref{eq:c1}) and (\ref{eq:c2}) without paying too much attention to definitions domains, one would write something like
\beq
``\Pi(p^2\in\mathds{R})=\hat\Pi(-p^2-i0^+)"\,,\label{eq:rep}
\eeq
from which one would conclude that the rule to get Minkowskian self-energies is to apply the replacement $P^2\to -(p^2+i0^+)$ in the Euclidean self-energies. However, this hides an important subtlety. 

In fact, combining Eqs.~(\ref{eq:c1}) and (\ref{eq:c2}) is only allowed when $\smash{p^2<0}$, in which case one actually retrieves Eq.~(\ref{eq:c0}).\footnote{The $i0^+$ can be ignored due to the analyticity of $\bar\Pi(z)$ accross the negative real axis.} But the replacement rule (\ref{eq:rep}) should not be extended too naively to the case where $\smash{p^2>0}$ because it takes the Euclidean self-energy out of its definition domain. This makes the replacement ambiguous as it leads to different functions depending on which expression of the Euclidean self-energy is used.\footnote{Take for instance $\ln P^2$ which, as long as $\smash{P^2>0}$, is identically equal to $1/2\ln (P^2)^2$. However, the replacement $\smash{P^2\to -(p^2+i0^+)}$ leads to two different functions of $p^2$: $\ln(-p^2-i0^+)$ and $1/2\ln(p^2+i0^+)^2$.} The actual recipe can be summarized as in Eq.~(\ref{eq:rep}), with, however, the important warning that the replacement cannot be done without first checking that the more general replacement $P^2\to -z$ in the chosen expression for the Euclidean self-energy leads to an analytic function of $z$ over $\mathds{C}/\mathds{R}_+$. In the case where it does not, the Euclidean expression for $P^2>0$ needs to be massaged until the replacement $P^2\to -z$ leads to an analytic function over $\mathds{C}/\mathds{R}_+$.\footnote{For instance, if we start from $1/2\ln (P^2)^2$, the replacement $\smash{P^2\to -z}$ leads to $1/2\ln (-z)^2$ which has a branchcut along the imaginary axis and is thus not analytic over $\mathds{C}/\mathds{R}^+$. However, if we consider instead $\ln P^2$ which equals $1/2\ln (P^2)^2$ as long as $\smash{P^2>0}$, the replacement $\smash{P^2\to -z}$ leads to $\ln (-z)$ whose branchcut is $\mathds{R}^+$.}

\subsection{Decomposition into master integrals}
Let us now prove the existence of $\bar\Pi(z)$. The first part of the proof  is based on decomposing the Minkowskian self-energies in terms of Minkowskian master integrals. Again, we could of course go through a similar algorithm than the one used to decompose the Euclidean self-energies in terms of Euclidean master integrals. But one can actually deduce the result in a very simple way from the result of the Euclidean decomposition (\ref{eq:reduc}). Indeed, except from the fact that Euclidean momenta $P$ become Minkowkian momenta $p$ with a different metric, $\delta_{\mu\nu}$ vs $g_{\mu\nu}$, and the fact that the square mass appears in the propagator with an extra minus sign in the Minkowskian case, it is a matter of tracking down extra factors of $i$ or $-i$. 

 At one-loop order, the Minkowskian diagrams involving two three-point vertices and two propagators bring an extra factor of $\smash{(-i)^4=1}$ with respect to the corresponding Euclidean diagrams, while the tadpole diagram brings an extra factor of $\smash{(-i)i=1}$. So, all these factors of $i$ cancel nicely. One should not forget though that there is an extra factor of $i$ between the Euclidean self-energies and the Minkowskian ones which are respectively $-1$ or $-i$ times the sum of  1PI diagrams. 
 
 Alltogether, this means that, in order to obtain the decomposition of the Minkowskian gluon self-energies in terms of Minkowskian master integrals, one can just take the decomposition (\ref{eq:reduc}) of the Euclidean self-energies in terms of Euclidean master integrals, and replace $m^2$ by $-m^2$ in the prefactors of the master integrals, while defining the Minkowskian master integrals $A_m$ and $B_{m_1m_2}(p^2)$ to include the extra factor of $i$. In fact, because the prefactor of $\hat B_{m_1m_2}(P^2)$ contains already an explicit factor of $m^2$, see Eq.~(\ref{eq:reduc}), which changes sign under $\smash{m^2\to -m^2}$, it is convenient to include a factor of $-i$ in the definition of  $B_{m_1m_2}(p^2)$.
 
 The rule is then to replace $P^2$ by $p^2$ and $m^2$ by $-m^2$ in the coefficients $\alpha_m(P^2/m^2)$ and $\beta_{m_1m_2}(P^2/m^2)$ of the decomposition. Then, the decomposition of the Minkowskian self-energies reads
\beq
\Pi(p^2) & = & \alpha_m\left(\frac{p^2}{-m^2}\right)A_m\nonumber\\
& + & m^2\!\!\!\!\sum_{m_i\in\{0,m\}}\!\!\!\!\beta_{m_1m_2}\left(\frac{p^2}{-m^2}\right)B_{m_1m_2}(p^2)\,,\label{eq:reduc2}
\eeq
with 
\beq
A_m & \equiv & i\int_q\frac{1}{q^2-m^2+i0^+}\,,\label{eq:AM}\\
B_{m_1m_2}(p^2) & \equiv & -i\int_q\frac{1}{q^2-m^2+i0^+}\frac{1}{(q+p)^2-m^2+i0^+}\,,\label{eq:BM}\nonumber\\
\eeq
and
\beq
\int_q\equiv \Lambda^{4-d}\int\frac{d^dq}{(2\pi)^d}\,.
\eeq
The coefficient functions $\alpha_m$ and $\beta_{m_1m_2}$ are exactly those of the corresponding Euclidean self-energies, see again App.~\ref{app:coeff}. 

Note that, because the coefficient functions depend only on the ratio $P^2/m^2$, the change $(P^2,m^2)\to (p^2,-m^2)$ is equivalent to $(P^2,m^2)\to (-p^2,m^2)$. We shall then retain the following rule of thumb: with the definitions (\ref{eq:AE})-(\ref{eq:BE}) and (\ref{eq:AM})-(\ref{eq:BM}) for the Euclidean and Minkowskian master integrals, the coefficients of the master decompositions of the corresponding self-energies are simply related by $\smash{P^2\to -p^2}$. 

\subsection{Connecting the master integrals}
The second step of our proof consists in unveiling the relation between the Euclidean and Minkowskian master integrals as defined above. In fact, in a certain sense to be clarified now, the rule of thumb of the previous section applies also to the master integrals. 

For instance, $\hat A_m$ and $A_m$ are momentum independent, and, as is well known, see the App.~\ref{app:B}, a Wick rotation leads to $\smash{\hat A_m=A_m}$.

The relation between $B_{m_1m_2}(p^2)$ and $\hat B_{m_1m_2}(P^2)$ is slightly more subtle. In the case where $\smash{p^2<0}$, a similar Wick rotation allows one to deduce that $B_{m_1m_2}(p^2<0)=\hat B_{m_1m_2}(-p^2)$, see App.~\ref{app:B}. The same caution as discussed above for the self-energies applies to the bubble integral, however: one cannot apply the replacement rule $\smash{P^2\to -p^2}$ too naively when $\smash{p^2>0}$.

In fact, in this case, the Wick rotation is not that useful. But it is possible to argue, see App.~\ref{app:B}, that there exists a unique analytic function $\bar B_{m_1m_2}(z)$ over $\mathds{C}/\mathds{R}_+$ which connects to the Euclidean bubble for $\smash{z=p^2<0}$ as
\beq
\bar B_{m_1m_2}(z=p^2<0)=\hat B_{m_1m_2}(-p^2)\,,\label{eq:eucl}
\eeq
and which gives the Minkowskian bubble for $\smash{z=p^2+i0^+}$ with $\smash{p^2>0}$ (and actually for any $\smash{p^2\in\mathds{R}}$):
\beq
\bar B_{m_1m_2}(z=p^2+i0^+)=B_{m_1m_2}(p^2)\,.\label{eq:cont}
\eeq
 So, once again, the Minkowskian bubble can be reached from the sole knowledge of the Euclidean bubble. As for the discussion of the self-energies above, the whole procedure can be summarized as that of applying the replacement $\smash{P^2\to -p^2-i0^+}$ in a given expression for the Euclidean bubble but only provided the more general replacement $\smash{P^2\to -z}$ in this same expression leads to an analytic function over $\mathds{C}/\mathds{R}_+$. If this is not the case, the Euclidean expression for $P^2>0$ needs to be massaged until the replacement $P^2\to -z$ leads to an analytic function over $\mathds{C}/\mathds{R}_+$.
 
 Coming back to the self-energies, let us now define
\beq
\bar\Pi(z) & = & \alpha_m\left(\frac{-z}{m^2}\right)A_m\nonumber\\
& + & m^2\!\!\!\!\sum_{m_i\in\{0,m\}}\!\!\!\!\beta_{m_1m_2}\left(\frac{-z}{m^2}\right)\bar B_{m_1m_2}(z)\,.
\eeq
This function is analytic over $\mathds{C}/\mathds{R}_+$ because both the coefficient functions $\alpha_m(-z/m^2)$ and $\beta_{m_1m_2}(-z/m^2)$, see App.~\ref{app:coeff}, as well as $\bar B_{m_1m_2}(z)$ are. Moreover, as it is trivially checked using Eqs.~(\ref{eq:reduc}) and (\ref{eq:eucl}), the so constructed $\bar\Pi(z)$ connects to the Euclidean self energy as in Eq.~(\ref{eq:c2}). We have thus shown the existence of the analytic self-energies anticipated in Sec.~\ref{sec:1loop_M}.

\subsection{Real-valued IR-safe renormalization factors}
Now that we know how to simply connect the Euclidean self-energies to the Minkowskian ones, let us discuss the renormalization of the Minkowskian propagators. As before, the one-loop renormalized propagators write
\beq
& & G_c^{-1}(p^2;q^2)=Z_c(q^2) p^2-\Pi_c(p^2)\,,\\
& & G_A^{-1}(p^2;q^2)\nonumber\\
& & \hspace{0.5cm}=\,Z_A(q^2)p^2-Z_A(q^2) Z_{m^2}(q^2) m^2-\Pi_A(p^2)\,,\nonumber\\
\eeq
where the renormalization scale is denoted as $q^2$. 

It would be tempting to define the IR-safe scheme as before. However, we should stress that the Minkowskian self-energies develop imaginary parts in the time-like region. If we stick for the moment to real-valued renormalization factors, the tree-level terms can only modify the real parts of the self-energies. For this reason, we propose the renormalization conditions
\beq
{\rm Re}\,G_c^{-1}(p^2=q^2;q^2) & = & q^2\,,\\
{\rm Re}\,G_A^{-1}(p^2=q^2;q^2) & = & q^2-m^2\,,
\eeq
where $q^2$ can be positive or negative. As before, we supplement  the above conditions with
\beq
Z_AZ_cZ_{m^2}=1 \quad {\rm and} \quad Z_AZ_c^2Z_{g^2}=1\,.\label{eq:nrt}
\eeq
All together, this defines the real-valued IR-safe scheme.

One peculiarity of this choice is that the space-like flows and the time-like flows are not connected to each other, see below. Therefore, it is not possible to associate a time-like flow to a given space-like flow. We shall later consider the possibility of using complex-valued renormalization factors and investigate whether this can address the issue, see Sec.~\ref{sec:C}. For the time being, we stick to the real-valued IR-safe scheme.

Proceeding as before, we find
\beq
Z_c(q^2) & = & 1+{\rm Re}\frac{\Pi_c(q^2)}{q^2}\,,\label{eq:Zcm}\\
Z_A(q^2) & = & 1-m^2{\rm Re}\frac{\Pi_c(q^2)}{(q^2)^2}+{\rm Re}\,\frac{\Pi_A(q^2)}{q^2}\,.\label{eq:ZAm}
\eeq
Note that these formulas encompass the Euclidean case as well. Indeed, using Eqs.~(\ref{eq:Zc})-(\ref{eq:ZA}) and (\ref{eq:c0}), and the fact that the Euclidean self-energies are real-valued, it is easily shown that
\beq
Z_X(q^2<0)=\hat Z_X(-q^2)\,.
\eeq 
From, this, we deduce that the flow for space-like $q^2$ is given by the Euclidean flow, as a function of $\smash{Q^2\equiv -q^2}$. 

In the time-like region, the connection can be done as we did for the self-energies. First, one defines 
\beq
\bar Z_c(z) & = & 1+\frac{\bar\Pi_c(z)}{z}\,,\label{eq:Zcz}\\
\bar Z_A(z) & = & 1-m^2\frac{\bar\Pi_c(z)}{z^2}+\frac{\bar\Pi_A(z)}{z}\,,\label{eq:ZAz}
\eeq
which are analytic over $\mathds{C}/\mathds{R}_+$ and connect to the Euclidean renormalization factors for $\smash{z=q^2<0}$ as
\beq
\bar Z_X(z=q^2<0)=\hat Z_X(-q^2)\,.
\eeq
Then one shows  that, for any real $q^2$, the Minkowskian renormalization factors are obtained from
\beq
Z_X(q^2\in\mathds{R})={\rm Re}\,\bar Z_X(q^2+i0^+)\,.\label{eq:prove}
\eeq
Indeed, upon evaluating $\bar Z_X(z)$ for $\smash{z=q^2+i0^+}$, and using Eq.~(\ref{eq:c1}), we find
\beq
\bar Z_c(q^2+i0^+) & = & 1+\frac{\Pi_c(q^2)}{q^2+i0^+}\,,\\
\bar Z_A(q^2+i0^+) & = & 1-m^2\frac{\Pi_c(q^2)}{(q^2+i0^+)^2}+\frac{\Pi_A(q^2)}{q^2+i0^+}\,.
\eeq
As we already discussed in the Euclidean case, the explicit poles in these expressions are spurious, so we can forget about the $i0^+$. Finally, taking the real parts, we arrive at
\beq
{\rm Re}\,\bar Z_c(q^2+i0^+) & = & 1+{\rm Re}\,\frac{\Pi_c(q^2)}{q^2}\,,\\
{\rm Re}\,\bar Z_A(q^+i0^+) & = & 1-m^2{\rm Re}\,\frac{\Pi_c(q^2)}{(q^2)^2}+{\rm Re}\,\frac{\Pi_A(q^2)}{q^2}\,,\nonumber\\
\eeq
whose right-hand sides are nothing but those of Eqs.~(\ref{eq:Zcm})-(\ref{eq:ZAm}), thus proving Eq.~(\ref{eq:prove}).

\begin{center}
\begin{figure}[t]
\includegraphics[height=0.3\textheight]{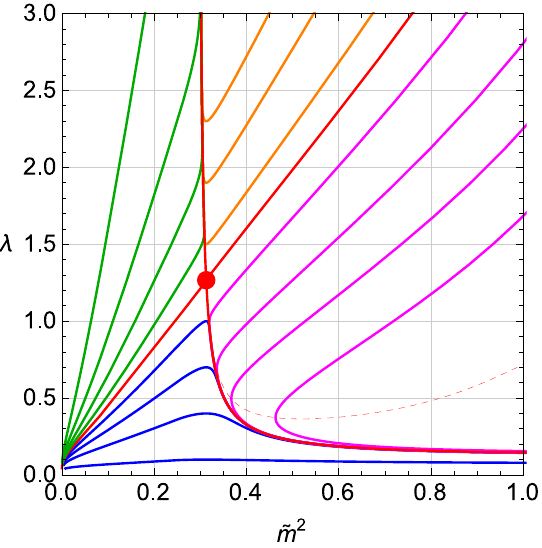}
\caption{
One-loop time-like Minkowskian flow of the Curci-Ferrari model in the real-valued IR-safe scheme, in the plane $(\tilde m^2,\lambda)$. The space-like flow is nothing but the mirror image of the Euclidean flow shown in Fig.~\ref{fig:diag}. The dashed lines correspond to points at which one of the two beta functions vanish. For the direction of the flow, see Fig.~\ref{fig:diag}.}
\label{fig:flow_m}
\end{figure}
\end{center}

\subsection{Anomalous dimensions and flow}
From the Minkowskian renormalization factors, we define
\beq
\gamma_X(q^2)=2q^2\frac{d\ln Z_X}{dq^2}\,.
\eeq
From the above considerations, we know that
\beq
\gamma_X(q^2)={\rm Re}\,\bar \gamma_X(q^2+i0^+)\,,\label{eq:dd}
\eeq
where
\beq
\bar\gamma_X(z)\equiv 2z\frac{d\ln \bar Z_X}{dz}
\eeq 
is the only analytic function over $\mathds{C}/\mathds{R}_+$ such that
\beq
\bar\gamma_X(z=q^2<0)=\hat\gamma_X(-q^2)\,.
\eeq
As before, the analytic anomalous dimensions $\bar\gamma_X(z)$ can be deduced from the Euclidean ones provided one uses Euclidean expressions such that the replacement $\smash{Q^2\to -z}$ leads to an analytic function over $\mathds{C}/\mathds{R}_+$. We have checked that this is the case if one starts from the expressions (\ref{eq:ggc})-(\ref{eq:gga}). According to Eq.~(\ref{eq:dd}), to get the Minkowskian anomalous dimensions in the considered scheme, we then just need to apply the replacement $T\equiv m^2/Q^2\to-m^2/q^2\equiv -t$ in those expressions, and take the real part. One then finds
\beq
\gamma_c(q^2) & = & -\frac{\lambda}{2t^2} {\rm Re}\,\Big[2t^2-2t +t^3 \ln(-t)\nonumber\\
& & \hspace{1.5cm} -\,(t-1)^2 (t+2) \ln (1-t)\Big]\,,
\eeq
and
\beq
\gamma_A(q^2) & = & \frac{\lambda}{6t^3} {\rm Re}\,\Bigg[-17 t^3-74t^2+12t+t^5 \ln (-t)\nonumber\\
& & \hspace{1.0cm}-\,(t+2)^2 (2t+3) (1-t)^2 \ln (1-t) \nonumber\\\nonumber\\
& & \hspace{1.0cm}-\,(-t)^{3\over2} \sqrt{4-t} \left(t^3+9t^2+20t+36\right)\nonumber\\
& & \hspace{1.5cm}\times\,\ln\!\left(\frac{\sqrt{4-t}-\sqrt{-t}}{\sqrt{4-t}+\sqrt{-t}}\right)\!\!\Bigg]\,.
\eeq
From the conditions (\ref{eq:nrt}) we have again that
\beq
\beta_{\lambda} & = & \lambda(\gamma_A+2\gamma_c)\,,\\
\beta_{\tilde m^2} & = & \tilde m^2(\gamma_A+\gamma_c-2)\,,
\eeq
with $\tilde m^2\equiv m^2/q^2=1/t$. 

\begin{center}
\begin{figure}[t]
\includegraphics[height=0.25\textheight]{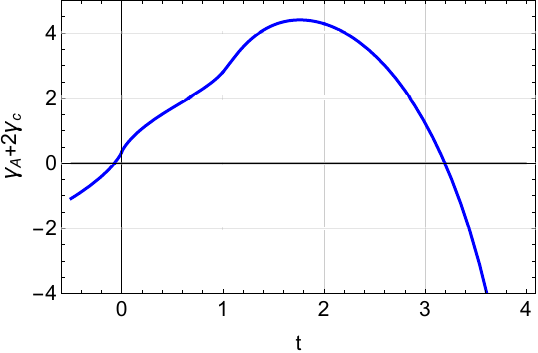}\qquad
\caption{Non-trivial zeros of $\beta_\lambda$ in the time-like ($t>0$) and space-like or Euclidean ($t<0$) domains. {\sf [[Show $\gamma_A+\gamma_c$]]}}
\label{fig:gac}
\end{figure}
\end{center}

The structure of the flow in the space-like region ($q^2<0$, or, equivalently $t<0$) is identical to that of the Euclidean flow upon the replacement $\smash{-t=T}$. This means that the flow diagram in this sector just looks like the mirror image of the one represented in Fig.~\ref{fig:diag} and we shall not reproduce it here. 

In the time-like region, the structure is similar, with four different types of theories, including theories that are asymptotically free both in the UV and in the IR, see Fig.~\ref{fig:flow_m}. We have of course an attractive UV fixed point at $(\tilde m^2,\lambda)=(0,0)$ but we also find a second non-trivial fixed point, obtained as before, that is by first noticing that the beta function of $\lambda$ write $\lambda^2 f(\tilde m^2)$ and by looking at the zeros of $f(\tilde m^2)$ for $\tilde m^2>0$. We find only one zero at $\tilde m^2=\tilde m^2_\star\simeq 0.31$. Then, by noticing that the beta function for $\tilde m^2$ reads $\tilde m^2(\lambda h(\tilde m^2)-2)$, we can make this function vanish at $\smash{\tilde m^2=\tilde m^2_\star}$ provided we choose $\smash{\lambda=\lambda_\star\equiv 2h(\tilde m^2_\star)\simeq 1.27}$, a not so large value of $\lambda$. In fact, the space-like and time-like non-trivial fixed-points are encompassed by the same function $f(\tilde m^2)$, see Fig.~\ref{fig:gac}. Note that because $\smash{\tilde m^2=0}$ is a flow trajectory, the space-like and time-like flows are not connected to each other in this scheme.

\begin{center}
\begin{figure}[t]
\includegraphics[height=0.3\textheight]{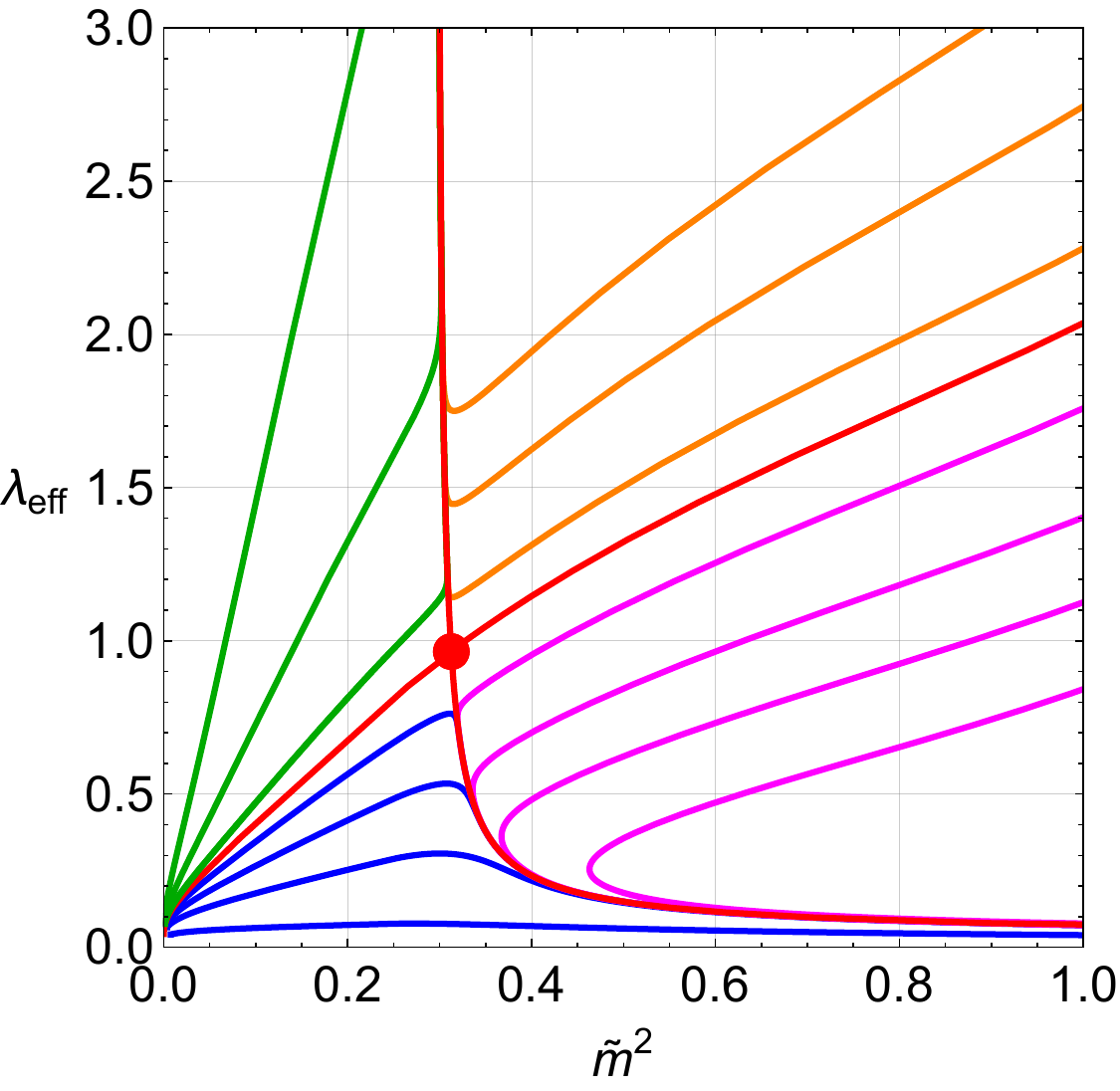}
\caption{
One-loop time-like Minkowskian flow of the Curci-Ferrari model in the real-valued IR-safe scheme in the plane $(\tilde m^2,\lambda_{\rm eff})$. The space-like flow is nothing but the Euclidean flow shown in Fig.~\ref{fig:diag}. For the direction of the flow, see Fig.~\ref{fig:diag}.}
\label{fig:flow_mtilde}
\end{figure}
\end{center}

\vglue-10mm

For completeness, we also show the time-like flow diagram in the $(\tilde m^2,\lambda_{\rm eff})$ plane in Fig.~\ref{fig:flow_mtilde}. The difference in the values of $\lambda$ at the space-like and time-like fixed points is drastically attenuated when expressed in terms of $\lambda_{\rm eff}$. We also notice that the value of $\lambda_{\rm eff}$ at the time-like fixed-point is $\simeq 0.97$, a value slightly below $1$.


\section{Analytic IR-safe flow?}\label{sec:C}
So far, we restricted to a conventional version of the IR-safe scheme involving real-valued renormalization factors. One inconvenience is that the space-like and time-like flows are not connected because the line $\smash{\tilde m^2=0}$ is a flow trajectory, thus preventing one from determining the relevant time-like trajectory knowing the space-like trajectory. It is thus interesting to look for other renormalization schemes that could allow to connect the two types of flow.

In this respect, it should be noticed that nothing prevents one from considering complex-valued renormalization factors. For instance, if one allows for complex-valued field renormalization factors $Z_c$ and $Z_A$, this just redefines what it is meant by renormalized correlation functions but should not affect scheme independent quantities such as poles of correlation functions or ratios of the same correlation function for different configurations of momenta.

Of course, in practice, one performs approximations and it is not true that scheme-independent quantities display exact scheme independence. For instance, a scheme independent quantity can feature a spurious dependence on the renormalization scale. This spurious dependence is usually turned into an asset as it can inform us about the quality of the approximation: good approximations should lead to small scheme dependences.  The same is true for schemes that use complex renormalization factors: real-valued scheme-independent quantities could display spurious imaginary parts but this can again be turned into a test of the quality of the approximation.

\begin{center}
\begin{figure}[t]
\includegraphics[height=0.25\textheight,width=0.45\textwidth]{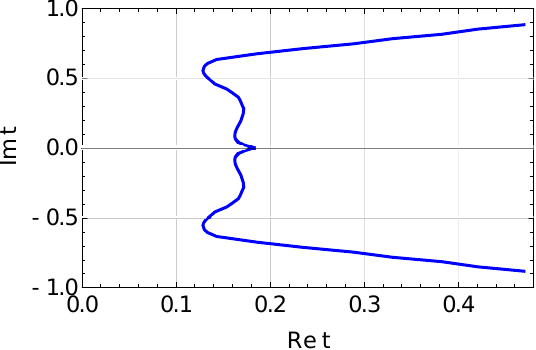}
\caption{Branchcut of the running parameters in the complex-valued IR-safe scheme, when integrating from $q_0^2=-1\,{\rm Gev}^2$ with $\tilde m^2_0=-(0.39)^2\,{\rm (GeV)^2}$ and $\lambda_0=0.26$. }
\label{fig:bc}
\end{figure}
\end{center}


\begin{center}
\begin{figure}[t]
\includegraphics[height=0.2\textheight]{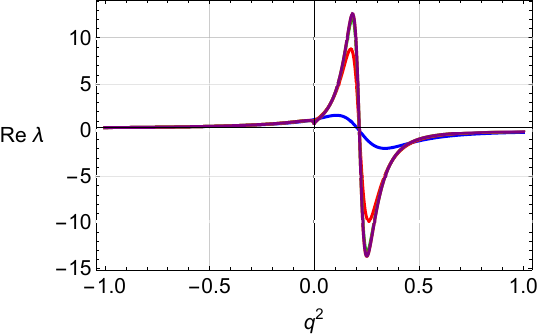}\\
\vglue4mm
\includegraphics[height=0.2\textheight]{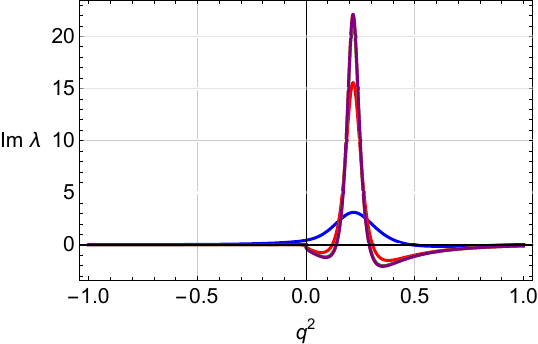}\\
\vglue4mm
\includegraphics[height=0.28\textheight]{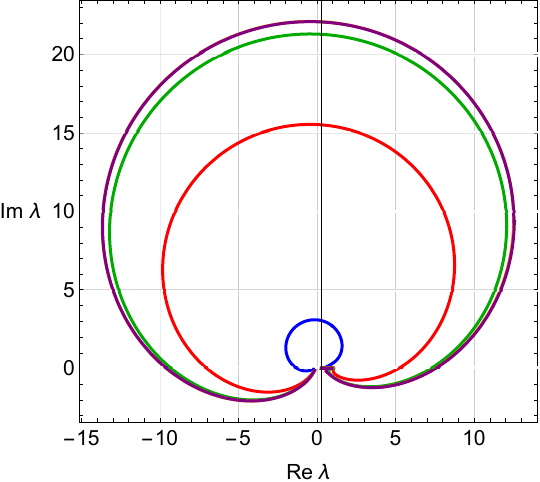}
\caption{Real and imaginary parts of the coupling from space-like to time-like running scales as one makes the $i0^+$ smaller and smaller: $10^{-1}$ (blue), $10^{-2}$ (red), $10^{-3}$ (green), $10^{-4}$ (orange), $10^{-5}$ (purple). The last two curves are essentially on top of each other.}
\label{fig:cross}
\end{figure}
\end{center}

\vglue-16mm

\begin{center}
\begin{figure}[t]
\includegraphics[height=0.2\textheight]{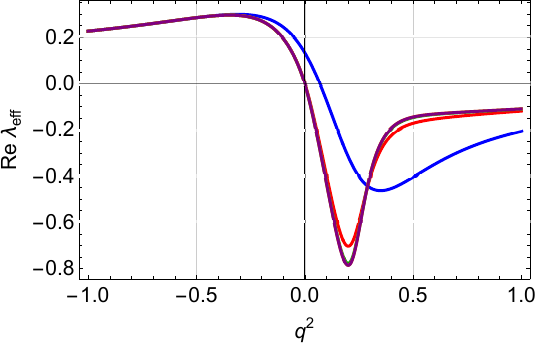}\\
\vglue4mm
\includegraphics[height=0.2\textheight]{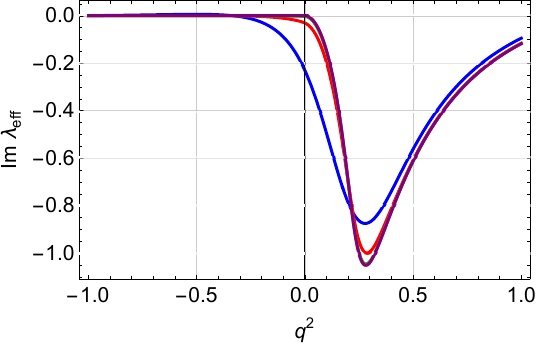}\\
\vglue4mm
\includegraphics[height=0.28\textheight]{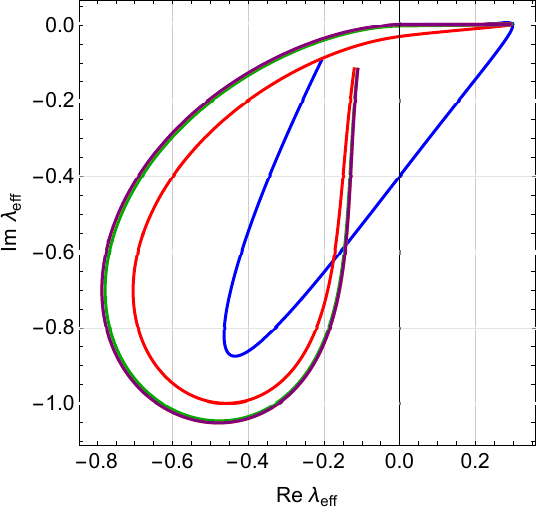}
\caption{Real and imaginary parts of the effective coupling $\lambda_{\rm eff}$ from space-like to time-like running scales as one makes the $i0^+$ smaller and smaller: $10^{-1}$ (blue), $10^{-2}$ (red), $10^{-3}$ (green), $10^{-4}$ (orange), $10^{-5}$ (purple). The last two curves are essentially on top of each other.}
\label{fig:cross2}
\end{figure}
\end{center}

\vglue-8mm

Having formulated these warnings, we can now consider a different version of the IR-safe renormalization scheme where, starting from the analytic inverse propagators
\beq
& & \bar G_c^{-1}(z;\omega)=\bar Z_c(\omega) z-\bar \Pi_c(z)\,,\\
& & \bar G_A^{-1}(z;\omega)\nonumber\\
& & \hspace{0.5cm}=\,\bar Z_A(\omega)z-\bar Z_A(\omega) \bar Z_{m^2}(\omega) m^2-\bar\Pi_A(z)\,,
\eeq
we impose the renormalization conditions, with $\omega$ an arbitrary complex number (note the absence of real parts),
\beq
\bar G_c^{-1}(z=\omega;\omega) & = & \omega\,,\\
\bar G_A^{-1}(z=\omega;\omega) & = & \omega-m^2\,,
\eeq
together with
\beq
\bar Z_A\bar Z_c\bar Z_{m^2}=1 \quad {\rm and} \quad \bar Z_A\bar Z_c^2\bar Z_{g^2}=1\,.
\eeq
It is a simple exercise to show that solving the constraints leads to $\bar Z_c$ and $\bar Z_A$ given in Eqs.~(\ref{eq:Zcz})-(\ref{eq:ZAz}), with $z$ replaced by $\omega$ of course. Note that, correspondingly, $\bar Z_{g^2}$ and $\bar Z_{m^2}$ become complex as well. This is not a problem provided we accept that the renormalized parameters $g^2$ and $m^2$ become complex while making sure that the bare parameters $\bar Z_{g^2}g^2$ and $\bar Z_{m^2}m^2$ remain real.

The flow of the parameters is now governed by the analytic anomalous dimensions (again, note the absence of real parts)
\beq
\bar\gamma_c(\omega) & = & -\frac{\lambda}{2t^2}\,\Big[2t^2-2t +t^3 \ln(-t)\nonumber\\
& & \hspace{1.5cm} -\,(t-1)^2 (t+2) \ln (1-t)\Big]\,,\label{eq:gca2}
\eeq
and
\beq
\bar\gamma_A(\omega) & = & \frac{\lambda}{6t^3} \Bigg[-17 t^3-74t^2+12t+t^5 \ln (-t)\nonumber\\
& & \hspace{1.0cm}-\,(t+2)^2 (2t+3) (1-t)^2 \ln (1-t) \nonumber\\\nonumber\\
& & \hspace{1.0cm}-\,(-t)^{3\over2} \sqrt{4-t} \left(t^3+9t^2+20t+36\right)\nonumber\\
& & \hspace{1.5cm}\times\,\ln\!\left(\frac{\sqrt{4-t}-\sqrt{-t}}{\sqrt{4-t}+\sqrt{-t}}\right)\!\!\Bigg]\,.\label{eq:gaa2}
\eeq
with $t=m^2/\omega\in\mathds{C}$, which enter the analytic beta functions as
\beq
\bar\beta_{\lambda} & = & \lambda(\bar\gamma_A+2\bar\gamma_c)\,,\\
\bar\beta_{\tilde m^2} & = & \tilde m^2(\bar\gamma_A+\bar\gamma_c-2)\,,
\eeq
with $\tilde m^2\equiv m^2/\omega=1/t$.

To solve the analytic flow, we start from a reference Euclidean point $q^2_0<0$ where the parameters $\tilde m^2_0<0$ and $\lambda_0$ have been fixed by comparison to Euclidean lattice data. Since the renormalization factors are real in the space-like region, the parameters should be taken real to make sure that the corresponding bare parameters are real as well. Then, one runs the analytic flow along a certain trajectory in the complex plane. As long as the complex $t$ that results from integrating the flow does not meet any branchcut of the beta functions, the flow is indeed analytic and independent of the chosen path. We find, however, that this analytic flow possesses a branchcut in the complex plane. For the Euclidean parameters $\tilde m^2_0=-(0.39)^2\,{\rm (GeV)^2}$ and $\lambda_0=0.26$ that best fit the lattice propagators for $q_0^2=-1\,{\rm Gev}^2$, the branchcut is represented in Fig.~\ref{fig:bc}.

We can nevertheless try to use this flow to cross from the space-like to the time-like region. We proceed as follows. We start in the space-like region with the values $\tilde m^2_0<0$ and $\lambda_0$ at a $q^2_0<0$, slightly displaced by a tiny imaginary part $i0^+$. Then, we integrate the analytic flow along the path $z=q^2+i0^+$. Whenever the flow enters the time-like region, the parameters develop imaginary parts. Eventually the flow crosses the branchcut in the time-like region. To keep our flow analytic, we go over the cut by adding or subtracting $2\pi i$ to the appropriate logaithms. We find that for a tiny displacement away from the real axis, only one of the logarithms needs to be updated.

\begin{center}
\begin{figure}[t]
\includegraphics[height=0.2\textheight]{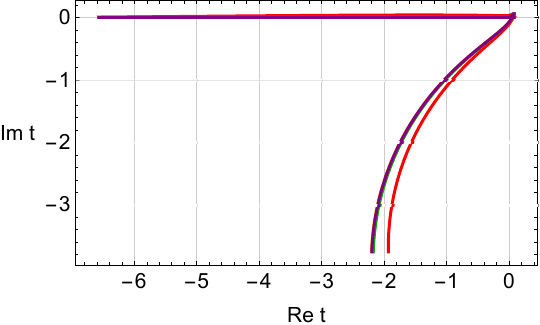}
\caption{Real and imaginary parts of $t$. The space-like flow runs from, almost real, very negative $t$ in the UV to very small and negative in the IR.}
\label{fig:cross3}
\end{figure}
\end{center}

\vglue-10mm

Our results for the real and imaginary parts of the coupling $\lambda$ are shown in Fig.~\ref{fig:cross}. We find that the coupling is not symmetric, as predicted in \cite{Milton:1998wi}, and takes much larger values (beyond the perturbative value $1$) in the time-like region. We note, however, that the results look much different when plotted in terms of the effective coupling $\lambda_{\rm eff}\equiv \lambda/(1-1/t)$, see Fig.~\ref{fig:cross2}. Although larger in the time-like region, the real and imaginary parts are found to be at most slightly above $1$ in modulus. For completeness, we also show in Fig.~\ref{fig:cross3}, the complex value of $t$ obtained upon integrating the flow.

\section{Conclusions}
We have extended the infrared-safe scheme of the Euclidean version of the Curci-Ferrari model to the Minkowskian version of the model, in such a way that the flow in the space-like region of the Minkowskian model coincides with the flow in the Euclidean model.

We have first used a scheme based on real-valued renormalization factors. In the time-like region, we have found a family of IR-safe trajectories whose coupling is bounded by a value slightly larger than $1$. Within this scheme, however, it is not possible to connect the time-like IR-safe trajectories to the corresponding space-like (or Euclidean) IR-safe trajectories. To cope with this, we have considered more general schemes based on complex-valued renormalization factors. We find that the time-like region can be connected with the space-like one. The parameter flow is implemented along a slightly imaginary trajectory, ensuring that the renormalization scale in the space-like region passes sufficiently close to the scale used to fix the parameters that best replicate the propagators obtained from Euclidean numerical simulations. This approach allows the Euclidean trajectory, which reproduces the propagators in the Euclidean region, to be extended into the time-like region. One subtle feature, however, is that it relies on the use of complex renormalized parameters. Although not forbidden, the latter could introduce, at any order of approximation, spurious imaginary parts in quantities that should be real were the theory to be solved exactly. It is thus important to assess how large are these spurious imaginary contributions in a given loop calculation, similar to the assessment of the spurious renormalization or scheme dependence of observables.

A natural continuation of this work is the study of the gluon and ghost two-point correlators in the time-like region. This we plan to do in a forthcoming article.

\begin{acknowledgments}
We would like to thank D.~M. van Egmond, G.~Krein, J.~Serreau, M.~Tissier and N.~Wschebor for fruitful discussions and useful suggestions on the manuscript.
\end{acknowledgments}


\appendix

\section{Euclidean decomposition into\\ master integrals}\label{app:coeff}
The decomposition of $\hat\Pi_c(P^2)/g^2N$ in terms of Euclidean master integrals yields the coefficients
\beq
\alpha_m(x) & = & \frac{x-1}{4}\,,\\
\beta_{00}(x) & = & \frac{x^2}{4}\,,\\
\beta_{0m}(x) & = & -\frac{(x+1)^2}{4}\,,\\
\beta_{m0}(x) & = & 0\,,\\
\beta_{mm}(x) & = & 0\,.
\eeq
while for $(d-2)\tilde\Pi_A(P^2)/g^2N$, these coefficient read
\beq
\alpha_m(x) & = & d^2+(x-3)d-\frac{7x-7+x^{-1}}{4}\,,\\
\beta_{00}(x) & = & \left(\frac{x}{2}-1\right)\frac{x}{4}\,,\\
\beta_{0m}(x) & = & (x+1)^2\left[d-\frac{x+6+x^{-1}}{4}\right],\\
\beta_{m0}(x) & = & 0\,,\\
\beta_{mm}(x) & = & (x+4)\left[\left(\frac{1}{2}-x\right)d+\frac{x^2+12-4}{8}\right],\nonumber\\
\eeq
see Ref.~\cite{Reinosa:2024vph} for a detailed explanation of the reduction procedure. We note that, because $\smash{\hat B_{m_1m_2}=\hat B_{m_2m_1}}$, the coefficients $\beta_{0m}$ and $\beta_{m0}$ are not fixed independently, but only their sum.

\section{Relating Euclidean and Minkowskian\\ master integrals}\label{app:B}
Let us start with the Minkowskian tapole $A_m$. Isolating the frequency integral, we have
\beq
A_m= i\int_{\vec{q}}\int_{-\infty}^{+\infty}\frac{dq_0}{2\pi}\frac{1}{q_0^2-\vec{q}^2-m^2+i0^+}\,.
\eeq
In the complex plane of the variable $q_0$, the poles are located at $\pm\sqrt{\vec{q}^2+m^2}\mp i0^+$. This means that the original real axis of integration can be Wick rotated anti-clockwise along the imaginary axis: $q_0\to i\tilde q_0$. We then find
\beq
A_m & = &  i\times i\int_{\vec{q}}\int_{-\infty}^{+\infty}\frac{d\tilde q_0}{2\pi}\frac{1}{-\tilde q_0^2-\vec{q}^2-m^2+i0^+}\nonumber\\
& = & \int_{\vec{q}}\int_{-\infty}^{+\infty}\frac{d\tilde q_0}{2\pi}\frac{1}{\tilde q_0^2+\vec{q}^2+m^2}\nonumber\\
& = & \int_Q \frac{1}{Q^2+m^2}=\hat A_m\,,
\eeq
where $Q=(\tilde q_0,\vec{q})$ is an Euclidean momentum and we neglected $0^+$ in the last steps as it becomes irrelevant after the Wick rotation has been performed. In summary, we have shown that, with our conventions, the tadpole Euclidean and Minkowskian master integrals agree.

Let us now consider the Minkowskian bubble. Using the Feynman trick, we have
\begin{widetext}
\beq
B_{m_1m_2}(p^2)=-i\int_0^1 dx \int_{\vec{q}} \int_{-\infty}^{+\infty}\frac{dq_0}{2\pi}\frac{1}{(q_0^2-\vec{q}^2-xm_1^2-(1-x)m_2^2+x(1-x)p^2+i0^+)^2}\,.\nonumber\\
\eeq
Let us first consider the case $p^2<0$ for it is simpler. In this case, we can write $\smash{p^2=-P^2}$ with $P^2>0$ and we are in a situation similar to that for the tadpole integral. Indeed, the poles are located at 
\beq
\underbrace{\pm\sqrt{\vec{q}^2+xm_1^2+(1-x)m_2^2+x(1-x)P^2}}_{\in\mathds{R}}\,\mp\, i0^+\,,
\eeq
and so one can perform the Wick rotation without problem:
\beq
& & B_{m_1m_2}(p^2=-P^2<0)\nonumber\\
& & \hspace{1.0cm}=\,-i\times i\int_0^1 dx \int_{\vec{q}} \int_{-\infty}^{+\infty}\frac{d\tilde q_0}{2\pi}\frac{1}{(-\tilde q_0^2-\vec{q}^2-xm_1^2-(1-x)m_2^2-x(1-x)P^2+i0^+)^2}\nonumber\\
& & \hspace{1.0cm}=\,\int_0^1 dx \int_{\vec{q}} \int_{-\infty}^{+\infty}\frac{d\tilde q_0}{2\pi}\frac{1}{(\tilde q_0^2+\vec{q}^2+xm_1^2+(1-x)m_2^2+x(1-x)P^2)^2}\nonumber\\
& & \hspace{1.0cm}=\,\int_Q \frac{1}{Q^2+m^2_1}\frac{1}{(Q+P)^2+m^2_2}=\hat B_{m_1m_2}(P^2)\,,
\eeq
\end{widetext}
where again $0^+$ was neglected in the last steps as it becomes irrelevant after the Wick rotation has been performed. In summary, in the case where $p^2<0$, we have found that
\beq
B_{m_1m_2}(p^2<0)=\hat B_{m_1m_2}(-p^2>0)\,.
\eeq
This means that the computation of the Minkowskian bubble for $\smash{p^2<0}$ boils down to that of the Euclidean bubble for $\smash{P^2=-p^2>0}$.

Let us now move to the case $p^2>0$. In this case, the Wick rotation is not that useful. But there still a simple way to deduce the Minkowskian bubble from the Euclidean one. Indeed, it can be shown, see below, that there exists a unique function $\bar B_{m_1m_2}(z\in\mathds C)$ analytic over $\mathds{C}/\mathds{R}^+$ which coincides with the Euclidean bubble for $z=p^2<0$ in the sense that:
\beq
\bar B_{m_1m_2}(z=p^2<0)=\hat B_{m_1m_2}(-p^2)\,,\label{eq:cond1}
\eeq
and which gives the Minkowskian bubble for $z=p^2+i0^+$, with $\smash{p^2\in\mathds{R}}$:
\beq
\bar B_{m_1m_2}(z=p^2+i0^+)=B_{m_1m_2}(p^2)\,.\label{eq:cond2}
\eeq
Thus from the sole knowledge of the Euclidean bubble and the requirement of analyticity over $\mathds{C}/\mathds{R}^+$, one can construct $\bar B_{m_1m_2}(z)$, from which one eventually obtains the Minkowskian bubble.

That $\bar B_{m_1m_2}(z)$ is unique follows from a well known theorem in complex analysis. Let us now check the existence by constructing $\bar B_{m_1m_2}(z)$ explicitely. One candidate (which will turn out not to be the good one) is

\begin{widetext}
\beq
\bar B_{m_1m_2}(z)\equiv-i\int_0^1 dx \int_{\vec{q}} \int_{-\infty}^{+\infty}\frac{dq_0}{2\pi}\frac{1}{(q_0^2-\vec{q}^2-xm_1^2-(1-x)m_2^2+x(1-x)z)^2}\,.\label{eq:try}
\eeq
It obeys condition (\ref{eq:cond2}) because
\beq
& & \bar B_{m_1m_2}(z=p^2+i0^+)\nonumber\\
& & \hspace{0.5cm}=\,-i\int_0^1 dx \int_{\vec{q}} \int_{-\infty}^{+\infty}\frac{dq_0}{2\pi}\frac{1}{(q_0^2-\vec{q}^2-xm_1^2-(1-x)m_2^2+x(1-x)(p^2+i0^+))^2}\nonumber\\
&  & \hspace{0.5cm}=\,-i\int_0^1 dx \int_{\vec{q}} \int_{-\infty}^{+\infty}\frac{dq_0}{2\pi}\frac{1}{(q_0^2-\vec{q}^2-xm_1^2-(1-x)m_2^2+x(1-x)p^2+i0^+)^2}\nonumber\\
& & \hspace{0.5cm}=\,B_{m_1m_2}(p^2)\,,
\eeq
\end{widetext}
where we have used that, without loss of generality, and because $x(1-x)>0$,  one can replace $x(1-x)z+i0^+$ by $x(1-x)(z+i0^+)$. 

It also looks as if the condition (\ref{eq:cond1}) was fulfilled. Indeed, using a Wick rotation, we saw above that
\beq
\bar B_{m_1m_2}(z=(p^2<0)+i0^+)=\hat B_{m_1m_2}(-p^2)\,.
\eeq
However, a similar reasonning shows that
\beq
\bar B_{m_1m_2}(z=(p^2<0)-i0^+)=-\hat B_{m_1m_2}(-p^2)\,,
\eeq
and thus $\bar B_{m_1m_2}(z)$ is not analytic across $\mathds{R}^-$. This can be fixed, however, by adding a factor ${\rm sgn}({\rm Im}\,z)$ in Eq.~(\ref{eq:try}). One can then argue that not only $\bar B_{m_1m_2}(z)$ is continuous across $\mathds{R}^-$ but that the same applies to its derivatives with respect to $z$. This ensures that $\bar B_{m_1m_2}(z)$ is analytic across $\mathds{R}^-$. Moreover, it now obeys the two conditions (\ref{eq:cond1}) and (\ref{eq:cond2}).

\section{Flow diagram}\label{app:diagram}

Let us here show how the particular structure of the beta functions allows one to deduce the main features of the flow diagram. We consider the Euclidean flow diagram as an example, see Fig.~\ref{fig:diag}, but exactly the same argumentation can be applied to the Minkowskian one.

Our analysis will be based on the fact that the beta functions have the general structure
\beq
\hat\beta_\lambda & = & \lambda^2f(\hat m^2),\\
\hat\beta_{\hat m^2} & = & \hat m^2\left(\lambda\,h(\hat m^2)-2\right).
\eeq
where the function $f(\hat m^2)$ has a unique zero $\hat m_\star^2$ along the positive real semi-axis, with $\smash{f(0)=-22/3}$ and $\smash{f(\infty)=1/3}$. For later purpose, we note that the function $h(\hat m^2)$ also vanishes at a lower value $\smash{\hat m^2_{\rm LP}<\hat m^2_\star}$, with $\smash{h(0)=-35/6}$ and $\smash{h(\infty)=1/3}$.

We can first notice that the beta function for $\lambda$ vanishes for $\smash{\lambda=0}$ while the beta function for $\hat m^2$ vanishes for $\smash{\hat m^2=0}$. This means that the lines $\smash{\lambda=0}$ and $\smash{\hat m^2=0}$ are both flow trajectories. At their intersection lies a fixed point (where both beta functions vanish) which is fully attractive in the UV, that is as $\smash{Q^2\to\infty}$. 

\begin{center}
\begin{figure}[t]
\includegraphics[height=0.3\textheight]{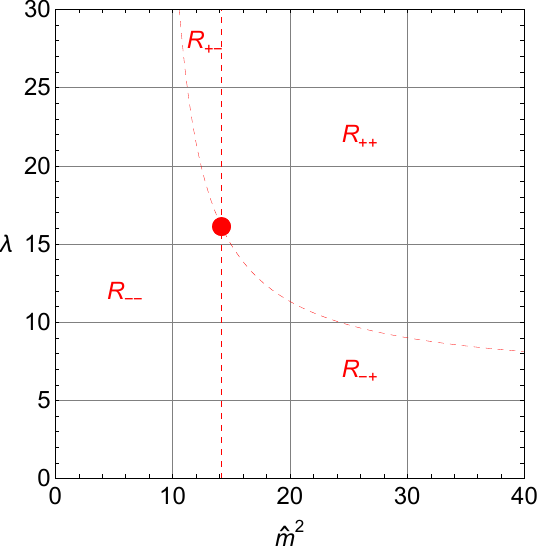}
\caption{}
\label{fig:regions}
\end{figure}
\end{center}

In addition, the beta function for $\lambda$ vanishes along the line $\smash{\hat m^2=\hat m^2_\star}$ while the beta function for $\hat m^2$ vanishes along the curve $\smash{\lambda(\hat m^2)=2/h(\hat m^2)}$, see the dashed curves in Fig.~\ref{fig:regions}. These curves meet at the point $\smash{(\hat m^2_\star,\hat\lambda_\star=2/h(\hat m^2_\star))}$ represented by a red dot in that figure. This is another fixed point of the flow which can be shown to have both attractive and repulsive IR directions. We note also that the curve $\smash{\lambda(\hat m^2)=2/h(\hat m^2)}$ has one asymptote at $\smash{\hat m^2=\hat m^2_{\rm LP}}$ and another one at $\smash{\lambda=\lambda_\infty\equiv 2/h(\infty)=6}$. 

\vglue-10mm

It is very convenient to subdivide the space of parameters into four regions, ${\cal R}_{--}$, ${\cal R}_{-+}$, ${\cal R}_{+-}$ and ${\cal R}_{++}$, delimited by the curves  $\smash{\hat m^2=\hat m^2_\star}$ and  $\smash{\lambda(\hat m^2)=2/h(\hat m^2)}$. Each region is characterized by definite signs of the beta functions $\beta_{\tilde m^2}$ and $\beta_\lambda$, as indicated by the indices $+$ or $-$. Then, as long as a given flow trajectory is in one region, $\lambda$ and $\hat m^2$ are monotonic functions of  the renormalization scale. Moreover, if the flow trajectory reaches a boundary separating two regions, one of the two parameters changes its monotonicity. The crossing to the other region happens with a slope $\smash{d\lambda/d\hat m^2=0}$ when crossing the line $\smash{\hat m^2=\hat m^2_\star}$ and with a slope $\smash{d\hat m^2/d\lambda=0}$ when crossing the curve $\smash{\lambda(\hat m^2)=2/h(\hat m^2)}$. 

From these remarks, together with the specific location that the regions ${\cal R}_{++}$, ${\cal R}_{+-}$, ${\cal R}_{-+}$ and ${\cal R}_{--}$ occupy in the plane\footnote{For instance, the region ${\cal R}_{--}$ contains $(0,0)$, \dots} $(\hat m^2,\lambda)$ it can be deduced that each trajectory crosses at most one of the curves $\smash{\hat m^2=\hat m^2_\star}$ and  $\smash{\lambda(\hat m^2)=2/h(\hat m^2)}$. Nothing excludes for the moment that certain trajectories do not cross any of these curves, specially because of the vertical asymptote of the curve $\smash{\lambda(\hat m^2)=2/h(\hat m^2)}$.

We can also deduce that, for each trajectory, only one of the two parameters is bounded from above or from below by the corresponding value at the non-trivial fixed-point. This defines four categories of trajectories:
\begin{itemize}
\item[$\bullet$] {\bf Type 1:} trajectories that connect ${\cal R}_{--}$ to ${\cal R}_{-+}$;
\item[$\bullet$] {\bf Type 2:} trajectories that remain in ${\cal R}_{--}$ or connect ${\cal R}_{--}$ to ${\cal R}_{+-}$;
\item[$\bullet$] {\bf Type 3:} trajectories that connect ${\cal R}_{++}$ to ${\cal R}_{-+}$;
\item[$\bullet$] {\bf Type 4:} trajectories that connect ${\cal R}_{++}$ to ${\cal R}_{+-}$.
\end{itemize}
Let us now study each type separately.

For trajectories of type $1$, the coupling decreases both in the UV and in the IR while $\hat m^2$ decreases in the UV and increases in the IR. The fate of the coupling can be studied by noticing that the flow of $1/\lambda$ is governed by the beta function
\beq
\hat\beta_{1/\lambda} & = & -f(\hat m^2)
\eeq
which does not depend on $\lambda$. Then, a formal integration leads to
\beq
\frac{1}{\lambda(\mu)}=\frac{1}{\lambda_0}-\frac{1}{2}\int_{\ln Q_0^2}^{\ln Q^2}d\ln (Q^2)' f(\hat m^2((Q^2)'))\,,\label{eq:integral}
\eeq
with $\lambda_0>0$. Since $\hat m^2$ decreases in the UV, we know that it converges towards a finite value. We do not know yet whether this value is $0$ but this does not matter because the value of $f(\hat m^2)$ in this limit is finite and strictly negative: $f_{--}^{\rm UV}<0$. Then, the integral in (\ref{eq:integral}) diverges negatively, so that $1/\lambda$ diverges positively as
\beq
\lambda(Q^2)\sim -\frac{2}{f_{--}^{\rm UV}}\frac{1}{\ln\frac{Q^2}{\bar Q^2}}\,,
\eeq
for $\smash{Q^2\to\infty}$. The same happens in the IR because $\hat m^2$ increases to a finite value or tends to $+\infty$ but in any case the value of $f(\hat m^2)$ in this limit is finite and strictly positive: $f_{-+}^{\rm IR}>0$. We then arrive at
\beq
\lambda(Q^2)\sim -\frac{2}{f_{-+}^{\rm IR}}\frac{1}{\ln\frac{Q^2}{\bar Q^2}}\,,
\eeq
for $\smash{Q^2\to 0}$.

To obtain the behavior of $\hat m^2$, we could similarly integrate its flow equation
\beq
\ln\frac{\hat m^2(Q^2)}{\hat m^2_0}=\frac{1}{2}\int_{\ln Q_0^2}^{\ln Q^2}d\ln (Q^2)' [\lambda\,h(\hat m^2)-2]\,.\label{eq:C7}
\eeq
Since the second term is subleading both in the UV and in the IR (since $\lambda$ goes to $0$ and $h(\hat m^2)$ remains finite), we deduce that
\beq
\ln\frac{\hat m^2(Q^2)}{\hat m^2_0}\sim\ln\frac{\bar Q^2}{Q^2}\,,\label{eq:C8}
\eeq
from which we can already conclude that $\hat m^2$ approaches $0$ in the UV and $+\infty$ in the IR, thus fixing the values of $f_{--}^{\rm UV}$ and $f_{+-}^{\rm IR}$ to 
\beq
f_{--}^{\rm UV}=f(0)=-\frac{22}{3} \quad {\rm and} \quad f_{-+}^{\rm IR}=f(\infty)=\frac{1}{3}\,,
\eeq
respectively. Notice that, from Eq.~(\ref{eq:C8}), it would be wrong to conclude that $\hat m^2(Q^2) Q^2$ becomes constant. This is because $\smash{u\sim v}$ does not always imply that $\smash{e^u\sim e^v}$. This is only true if we have in addition $\smash{u-v\to 0}$. 

Trajectories of type 2 have the same UV behavior as type 1. In the IR in contrast, because $f_{--}^{\rm IR}<0$ or $f_{+-}^{\rm IR}<0$ (we need to treat both cases because some of the trajectories could remain in ${\cal R}_{--}$ while others may cross to ${\cal R}_{+-}$), we deduce that for any starting positive $\lambda_0$, there is a finite scale $Q^2_{\rm LP}<Q_0^2$ at which $1/\lambda$ vanishes and thus $\lambda$ diverges. This is an infrared Landau pole and, in its vicinity, one can write
\beq
\lambda(Q^2)\sim -\frac{2}{f_{\pm-}^{\rm IR}}\frac{1}{\ln\frac{Q^2}{Q^2_{\rm LP}}}\,,
\eeq
Plugging this back  in Eq.~(\ref{eq:C7}), we find that the only possibility is that $\hat m^2$ approches $\hat m^2_{\rm LP}$, the zero of $h(\hat m^2)$, at the Landau pole. Thus, irrespectively of whether they remain in region ${\cal R}_{--}$ or they cross to ${\cal R}_{-+}$, the trajectories of type 2 approach the asymptote $\smash{\hat m^2=\hat m^2_{\rm LP}}$.

Trajectories of type 3 are similar to type 2 but with UV and IR reversed. In this case, there are no asymptotes and $\hat m^2$ approaches $\infty$ at the Landau pole. Finally, trajectories of the type 4 have both UV and IR Landau poles, with $\hat m^2$ approaching $\hat m^2_{\rm LP}$ in the infrared and $\infty$ in the ultraviolet.

From these considerations, we conclude that, the flow diagram of Fig.~\ref{fig:diag} describes four types of trajectories (and, therefore, four types of theories):
\begin{itemize}

\item[$\bullet$] {\bf Type 1:} theories that are asymptotically free both in the UV and in the IR (blue trajectories in Fig.~\ref{fig:diag}). For these theories, the parameter $\hat m^2$ goes to $0$ in the UV and to $\infty$ in the IR;

\item[$\bullet$] {\bf Type 2:} theories that are asymptotically free in the UV and possess an IR Landau pole (green trajectories). For these theories, the parameter $\hat m^2$ goes to $0$ in the UV and to a finite value $\hat m^2_{\rm LP}$ as one approaches the IR Landau pole;

\item[$\bullet$] {\bf Type 3:} theories that are asymptotically free in the IR and possess an UV Landau pole (purple trajectories). For these theories, the parameter $\hat m^2$ goes to infinity in the IR and also as one approaches the UV Landau pole;

\item[$\bullet$] {\bf Type 4:} theories that possess both an IR and an UV Landau pole (orange trajectories). For these theories, the parameter $\hat m^2$ approaches $\hat m^2_{\rm LP}$ as one approches the IR Landau pole, and diverges as one approaches the UV Landau pole.

\end{itemize}

\bibliographystyle{unsrt}
\bibliography{references}

\begin{thebibliography}{10}

\bibitem{Tissier:2011ey}
Matthieu Tissier and Nicolas Wschebor.
\newblock {An Infrared Safe perturbative approach to Yang-Mills correlators}.
\newblock {\em Phys. Rev. D}, 84:045018, 2011.

\bibitem{Curci:1976bt}
G.~Curci and R.~Ferrari.
\newblock {On a Class of Lagrangian Models for Massive and Massless Yang-Mills
  Fields}.
\newblock {\em Nuovo Cim. A}, 32:151--168, 1976.

\bibitem{Tissier:2010ts}
Matthieu Tissier and Nicolas Wschebor.
\newblock {Infrared propagators of Yang-Mills theory from perturbation theory}.
\newblock {\em Phys. Rev. D}, 82:101701, 2010.

\bibitem{Pelaez:2013cpa}
Marcela Pelaez, Matthieu Tissier, and Nicolas Wschebor.
\newblock {Three-point correlation functions in Yang-Mills theory}.
\newblock {\em Phys. Rev. D}, 88:125003, 2013.

\bibitem{Pelaez:2014mxa}
M.~Pel\'aez, M.~Tissier, and N.~Wschebor.
\newblock {Two-point correlation functions of QCD in the Landau gauge}.
\newblock {\em Phys. Rev. D}, 90:065031, 2014.

\bibitem{Pelaez:2015tba}
Marcela Pel\'aez, Matthieu Tissier, and Nicol\'as Wschebor.
\newblock {Quark-gluon vertex from the Landau gauge Curci-Ferrari model}.
\newblock {\em Phys. Rev. D}, 92(4):045012, 2015.

\bibitem{Figueroa:2021sjm}
Felipe Figueroa and Marcela Pel\'aez.
\newblock {One-loop unquenched three-gluon and ghost-gluon vertices in the
  Curci-Ferrari model}.
\newblock {\em Phys. Rev. D}, 105(9):094005, 2022.

\bibitem{Gracey:2019xom}
John~A. Gracey, Marcela Pel\'aez, Urko Reinosa, and Matthieu Tissier.
\newblock {Two loop calculation of Yang-Mills propagators in the Curci-Ferrari
  model}.
\newblock {\em Phys. Rev. D}, 100(3):034023, 2019.

\bibitem{Barrios:2020ubx}
Nahuel Barrios, Marcela Pel\'aez, Urko Reinosa, and Nicol\'as Wschebor.
\newblock {The ghost-antighost-gluon vertex from the Curci-Ferrari model:
  Two-loop corrections}.
\newblock {\em Phys. Rev. D}, 102:114016, 2020.

\bibitem{Barrios:2021cks}
Nahuel Barrios, John~A. Gracey, Marcela Pel\'aez, and Urko Reinosa.
\newblock {Two-loop corrections to the QCD propagators within the Curci-Ferrari
  model}.
\newblock {\em Phys. Rev. D}, 104(9):094019, 2021.

\bibitem{Barrios:2022hzr}
Nahuel Barrios, Marcela Pel\'aez, and Urko Reinosa.
\newblock {Two-loop three-gluon vertex from the Curci-Ferrari model and its
  leading infrared behavior to all loop orders}.
\newblock {\em Phys. Rev. D}, 106(11):114039, 2022.

\bibitem{Barrios:2024ixj}
Nahuel Barrios, Philipe De~Fabritiis, and Marcela Pel\'aez.
\newblock {Four-gluon vertex from the Curci-Ferrari model at one-loop order}.
\newblock {\em Phys. Rev. D}, 109(9):L091502, 2024.

\bibitem{Reinosa:2014zta}
U.~Reinosa, J.~Serreau, M.~Tissier, and N.~Wschebor.
\newblock {Deconfinement transition in SU(2) Yang-Mills theory: A two-loop
  study}.
\newblock {\em Phys. Rev. D}, 91:045035, 2015.

\bibitem{Reinosa:2015gxn}
U.~Reinosa, J.~Serreau, M.~Tissier, and N.~Wschebor.
\newblock {Two-loop study of the deconfinement transition in Yang-Mills
  theories: SU(3) and beyond}.
\newblock {\em Phys. Rev. D}, 93(10):105002, 2016.

\bibitem{Reinosa:2016xaj}
Urko Reinosa.
\newblock {A perturbative approach to the confinement-deconfinement phase
  transition}.
\newblock {\em EPJ Web Conf.}, 129:00032, 2016.

\bibitem{Maelger:2018vow}
Jan Maelger, Urko Reinosa, and Julien Serreau.
\newblock {Universal aspects of the phase diagram of QCD with heavy quarks}.
\newblock {\em Phys. Rev. D}, 98(9):094020, 2018.

\bibitem{Surkau:2024zfb}
Victor Tomas~Mari Surkau and Urko Reinosa.
\newblock {Deconfinement transition within the Curci-Ferrari model:
  Renormalization scale and scheme dependences}.
\newblock {\em Phys. Rev. D}, 109(9):094033, 2024.

\bibitem{Pelaez:2017bhh}
Marcela Pel\'aez, Urko Reinosa, Julien Serreau, Matthieu Tissier, and Nicol\'as
  Wschebor.
\newblock {Small parameters in infrared quantum chromodynamics}.
\newblock {\em Phys. Rev. D}, 96(11):114011, 2017.

\bibitem{Pelaez:2020ups}
Marcela Pel\'aez, Urko Reinosa, Julien Serreau, Matthieu Tissier, and Nicol\'as
  Wschebor.
\newblock {Spontaneous chiral symmetry breaking in the massive Landau gauge:
  realistic running coupling}.
\newblock {\em Phys. Rev. D}, 103(9):094035, 2021.

\bibitem{Pelaez:2022rwx}
Marcela Pel\'aez, Urko Reinosa, Julien Serreau, and Nicol\'as Wschebor.
\newblock {Small parameters in infrared QCD: The pion decay constant}.
\newblock {\em Phys. Rev. D}, 107(5):054025, 2023.

\bibitem{Pelaez:2024mtq}
Marcela Pel\'aez, Urko Reinosa, Julien Serreau, Matthieu Tissier, and Nicol\'as
  Wschebor.
\newblock {Quark confinement from an infrared safe approach}.
\newblock 5 2024.

\bibitem{Roberts:1994dr}
Craig~D. Roberts and Anthony~G. Williams.
\newblock {Dyson-Schwinger equations and their application to hadronic
  physics}.
\newblock {\em Prog. Part. Nucl. Phys.}, 33:477--575, 1994.

\bibitem{Alkofer:2000wg}
Reinhard Alkofer and Lorenz von Smekal.
\newblock {The Infrared behavior of QCD Green's functions: Confinement
  dynamical symmetry breaking, and hadrons as relativistic bound states}.
\newblock {\em Phys. Rept.}, 353:281, 2001.

\bibitem{Fischer:2006ub}
Christian~S. Fischer.
\newblock {Infrared properties of QCD from Dyson-Schwinger equations}.
\newblock {\em J. Phys. G}, 32:R253--R291, 2006.

\bibitem{Roberts:2007ji}
C.~D. Roberts.
\newblock {Hadron Properties and Dyson-Schwinger Equations}.
\newblock {\em Prog. Part. Nucl. Phys.}, 61:50--65, 2008.

\bibitem{Binosi:2009qm}
Daniele Binosi and Joannis Papavassiliou.
\newblock {Pinch Technique: Theory and Applications}.
\newblock {\em Phys. Rept.}, 479:1--152, 2009.

\bibitem{Cloet:2013jya}
Ian~C. Cloet and Craig~D. Roberts.
\newblock {Explanation and Prediction of Observables using Continuum Strong
  QCD}.
\newblock {\em Prog. Part. Nucl. Phys.}, 77:1--69, 2014.

\bibitem{Aguilar:2015bud}
A.~C. Aguilar, D.~Binosi, and J.~Papavassiliou.
\newblock {The Gluon Mass Generation Mechanism: A Concise Primer}.
\newblock {\em Front. Phys. (Beijing)}, 11(2):111203, 2016.

\bibitem{Huber:2018ned}
Markus~Q. Huber.
\newblock {Nonperturbative properties of Yang\textendash{}Mills theories}.
\newblock {\em Phys. Rept.}, 879:1--92, 2020.

\bibitem{Papavassiliou:2022wrb}
J.~Papavassiliou.
\newblock {Emergence of mass in the gauge sector of QCD*}.
\newblock {\em Chin. Phys. C}, 46(11):112001, 2022.

\bibitem{Binosi:2014aea}
Daniele Binosi, Lei Chang, Joannis Papavassiliou, and Craig~D. Roberts.
\newblock {Bridging a gap between continuum-QCD and ab initio predictions of
  hadron observables}.
\newblock {\em Phys. Lett. B}, 742:183--188, 2015.

\bibitem{Maris:1997tm}
Pieter Maris and Craig~D. Roberts.
\newblock {Pi- and K meson Bethe-Salpeter amplitudes}.
\newblock {\em Phys. Rev. C}, 56:3369--3383, 1997.

\bibitem{Maris:2003vk}
Pieter Maris and Craig~D. Roberts.
\newblock {Dyson-Schwinger equations: A Tool for hadron physics}.
\newblock {\em Int. J. Mod. Phys. E}, 12:297--365, 2003.

\bibitem{Eichmann:2008ef}
G.~Eichmann, I.~C. Cloet, R.~Alkofer, A.~Krassnigg, and C.~D. Roberts.
\newblock {Toward unifying the description of meson and baryon properties}.
\newblock {\em Phys. Rev. C}, 79:012202, 2009.

\bibitem{Cloet:2008re}
I.~C. Cloet, G.~Eichmann, B.~El-Bennich, T.~Klahn, and C.~D. Roberts.
\newblock {Survey of nucleon electromagnetic form factors}.
\newblock {\em Few Body Syst.}, 46:1--36, 2009.

\bibitem{Boucaud:2008ky}
Philippe Boucaud, J.~P. Leroy, A.~Le~Yaouanc, J.~Micheli, O.~Pene, and
  J.~Rodriguez-Quintero.
\newblock {On the IR behaviour of the Landau-gauge ghost propagator}.
\newblock {\em JHEP}, 06:099, 2008.

\bibitem{Eichmann:2009qa}
G.~Eichmann, R.~Alkofer, A.~Krassnigg, and D.~Nicmorus.
\newblock {Nucleon mass from a covariant three-quark Faddeev equation}.
\newblock {\em Phys. Rev. Lett.}, 104:201601, 2010.

\bibitem{Fischer:2008uz}
Christian~S. Fischer, Axel Maas, and Jan~M. Pawlowski.
\newblock {On the infrared behavior of Landau gauge Yang-Mills theory}.
\newblock {\em Annals Phys.}, 324:2408--2437, 2009.

\bibitem{Rodriguez-Quintero:2010qad}
J.~Rodriguez-Quintero.
\newblock {On the massive gluon propagator, the PT-BFM scheme and the
  low-momentum behaviour of decoupling and scaling DSE solutions}.
\newblock {\em JHEP}, 01:105, 2011.

\bibitem{Pennington:2011xs}
M.~R. Pennington and D.~J. Wilson.
\newblock {Are the Dressed Gluon and Ghost Propagators in the Landau Gauge
  presently determined in the confinement regime of QCD?}
\newblock {\em Phys. Rev. D}, 84:119901, 2011.

\bibitem{Huber:2012zj}
Markus~Q. Huber, Axel Maas, and Lorenz von Smekal.
\newblock {Two- and three-point functions in two-dimensional Landau-gauge
  Yang-Mills theory: Continuum results}.
\newblock {\em JHEP}, 11:035, 2012.

\bibitem{Roberts:2020hiw}
Craig~D Roberts.
\newblock {Empirical Consequences of Emergent Mass}.
\newblock {\em Symmetry}, 12(9):1468, 2020.

\bibitem{Gao:2021wun}
Fei Gao, Joannis Papavassiliou, and Jan~M. Pawlowski.
\newblock {Fully coupled functional equations for the quark sector of QCD}.
\newblock {\em Phys. Rev. D}, 103(9):094013, 2021.

\bibitem{Huber:2016tvc}
Markus~Q. Huber.
\newblock {Correlation functions of three-dimensional Yang-Mills theory from
  Dyson-Schwinger equations}.
\newblock {\em Phys. Rev. D}, 93(8):085033, 2016.

\bibitem{Huber:2020keu}
Markus~Q. Huber.
\newblock {Correlation functions of Landau gauge Yang-Mills theory}.
\newblock {\em Phys. Rev. D}, 101:114009, 2020.

\bibitem{Aguilar:2023mam}
A.~C. Aguilar, M.~.~N. Ferreira, D.~Iba\~nez, and J.~Papavassiliou.
\newblock {Schwinger displacement of the quark\textendash{}gluon vertex}.
\newblock {\em Eur. Phys. J. C}, 83(10):967, 2023.

\bibitem{Aguilar:2023qqd}
A.~C. Aguilar, M.~N. Ferreira, J.~Papavassiliou, and L.~R. Santos.
\newblock {Planar degeneracy of the three-gluon vertex}.
\newblock {\em Eur. Phys. J. C}, 83(6):549, 2023.

\bibitem{Aguilar:2024fen}
A.~C. Aguilar, M.~N. Ferreira, J.~Papavassiliou, and L.~R. Santos.
\newblock {Four-gluon vertex in collinear kinematics}.
\newblock {\em Eur. Phys. J. C}, 84(7):676, 2024.

\bibitem{Cyrol:2014kca}
Anton~K. Cyrol, Markus~Q. Huber, and Lorenz von Smekal.
\newblock {A Dyson\textendash{}Schwinger study of the four-gluon vertex}.
\newblock {\em Eur. Phys. J. C}, 75:102, 2015.

\bibitem{Braun:2007bx}
Jens Braun, Holger Gies, and Jan~M. Pawlowski.
\newblock {Quark Confinement from Color Confinement}.
\newblock {\em Phys. Lett. B}, 684:262--267, 2010.

\bibitem{Fister:2013bh}
Leonard Fister and Jan~M. Pawlowski.
\newblock {Confinement from Correlation Functions}.
\newblock {\em Phys. Rev. D}, 88:045010, 2013.

\bibitem{Pawlowski:2003hq}
Jan~M. Pawlowski, Daniel~F. Litim, Sergei Nedelko, and Lorenz von Smekal.
\newblock {Infrared behavior and fixed points in Landau gauge QCD}.
\newblock {\em Phys. Rev. Lett.}, 93:152002, 2004.

\bibitem{Pawlowski:2005xe}
Jan~M. Pawlowski.
\newblock {Aspects of the functional renormalisation group}.
\newblock {\em Annals Phys.}, 322:2831--2915, 2007.

\bibitem{Cyrol:2017ewj}
Anton~K. Cyrol, Mario Mitter, Jan~M. Pawlowski, and Nils Strodthoff.
\newblock {Nonperturbative quark, gluon, and meson correlators of unquenched
  QCD}.
\newblock {\em Phys. Rev. D}, 97(5):054006, 2018.

\bibitem{Cyrol:2018xeq}
Anton~K. Cyrol, Jan~M. Pawlowski, Alexander Rothkopf, and Nicolas Wink.
\newblock {Reconstructing the gluon}.
\newblock {\em SciPost Phys.}, 5(6):065, 2018.

\bibitem{Corell:2018yil}
Lukas Corell, Anton~K. Cyrol, Mario Mitter, Jan~M. Pawlowski, and Nils
  Strodthoff.
\newblock {Correlation functions of three-dimensional Yang-Mills theory from
  the FRG}.
\newblock {\em SciPost Phys.}, 5(6):066, 2018.

\bibitem{Blaizot:2021ikl}
Jean-Paul Blaizot, Jan~M. Pawlowski, and Urko Reinosa.
\newblock {Functional renormalization group and 2PI effective action
  formalism}.
\newblock {\em Annals Phys.}, 431:168549, 2021.

\bibitem{Horak:2021pfr}
Jan Horak, Joannis Papavassiliou, Jan~M. Pawlowski, and Nicolas Wink.
\newblock {Ghost spectral function from the spectral Dyson-Schwinger equation}.
\newblock {\em Phys. Rev. D}, 104, 2021.

\bibitem{Reinosa:2017qtf}
Urko Reinosa, Julien Serreau, Matthieu Tissier, and Nicol\'as Wschebor.
\newblock {How nonperturbative is the infrared regime of Landau gauge
  Yang-Mills correlators?}
\newblock {\em Phys. Rev. D}, 96(1):014005, 2017.

\bibitem{Siringo:2015wtx}
Fabio Siringo.
\newblock {Analytical study of Yang\textendash{}Mills theory in the infrared
  from first principles}.
\newblock {\em Nucl. Phys. B}, 907:572--596, 2016.

\bibitem{Siringo:2019qwx}
Fabio Siringo.
\newblock {Calculation of the nonperturbative strong coupling from first
  principles}.
\newblock {\em Phys. Rev. D}, 100(7):074014, 2019.

\bibitem{Comitini:2020ozt}
Giorgio Comitini and Fabio Siringo.
\newblock {One-loop RG improvement of the screened massive expansion in the
  Landau gauge}.
\newblock {\em Phys. Rev. D}, 102(9):094002, 2020.

\bibitem{Comitini:2021kxj}
Giorgio Comitini, Daniele Rizzo, Massimiliano Battello, and Fabio Siringo.
\newblock {Screened massive expansion of the quark propagator in the Landau
  gauge}.
\newblock {\em Phys. Rev. D}, 104(7):074020, 2021.

\bibitem{Mintz:2017qri}
B.~W. Mintz, L.~F. Palhares, S.~P. Sorella, and A.~D. Pereira.
\newblock {Ghost-gluon vertex in the presence of the Gribov horizon}.
\newblock {\em Phys. Rev. D}, 97(3):034020, 2018.

\bibitem{Barrios:2024idr}
Nahuel Barrios, Marcela Pel\'aez, Marcelo~S. Guimaraes, Bruno~W. Mintz, and
  Let\'\i{}cia~F. Palhares.
\newblock {Ghost-gluon vertex in the presence of the Gribov horizon: General
  kinematics}.
\newblock {\em Phys. Rev. D}, 109(9):094039, 2024.

\bibitem{deBrito:2024ffa}
Gustavo~P. de~Brito and Antonio~D. Pereira.
\newblock {Infrared gluon propagator in the refined Gribov-Zwanziger scenario
  at one-loop order in the Landau gauge}.
\newblock {\em Phys. Rev. D}, 110(7):074005, 2024.

\bibitem{Hayashi:2020few}
Yui Hayashi and Kei-Ichi Kondo.
\newblock {Complex poles and spectral functions of Landau gauge QCD and
  QCD-like theories}.
\newblock {\em Phys. Rev. D}, 101(7):074044, 2020.

\bibitem{Passarino:1978jh}
G.~Passarino and M.~J.~G. Veltman.
\newblock {One Loop Corrections for e+ e- Annihilation Into mu+ mu- in the
  Weinberg Model}.
\newblock {\em Nucl. Phys. B}, 160:151--207, 1979.

\bibitem{Gracey:2002yt}
J.~A. Gracey.
\newblock {Three loop MS-bar renormalization of the Curci-Ferrari model and the
  dimension two BRST invariant composite operator in QCD}.
\newblock {\em Phys. Lett. B}, 552:101--110, 2003.

\bibitem{Dudal:2002pq}
D.~Dudal, H.~Verschelde, and S.~P. Sorella.
\newblock {The Anomalous dimension of the composite operator A**2 in the Landau
  gauge}.
\newblock {\em Phys. Lett. B}, 555:126--131, 2003.

\bibitem{Wschebor:2007vh}
Nicolas Wschebor.
\newblock {Some non-renormalization theorems in Curci-Ferrari model}.
\newblock {\em Int. J. Mod. Phys. A}, 23:2961--2973, 2008.

\bibitem{Tissier:2008nw}
Matthieu Tissier and Nicolas Wschebor.
\newblock {Gauged supersymmetries in Yang-Mills theory}.
\newblock {\em Phys. Rev. D}, 79:065008, 2009.

\bibitem{Taylor:1971ff}
J.~C. Taylor.
\newblock {Ward Identities and Charge Renormalization of the Yang-Mills Field}.
\newblock {\em Nucl. Phys. B}, 33:436--444, 1971.

\bibitem{Reinosa:2024vph}
Urko Reinosa.
\newblock {Three lectures on the Curci-Ferrari model}.
\newblock In {\em {3rd Joint ICTP-SAIFR/ICTP-Trieste Summer School on Particle
  Physics}}, 3 2024.

\bibitem{Milton:1998wi}
K.~A. Milton and I.~L. Solovtsov.
\newblock {Can the QCD effective charge be symmetrical in the Euclidean and the
  Minkowskian regions?}
\newblock {\em Phys. Rev. D}, 59:107701, 1999.

\end{thebibliography}

\end{document}